\documentclass[aps,prd,twocolumn,tightenlines,superscriptaddress,byrevtex]{revtex4}

\usepackage{graphicx}
\usepackage{amsmath}

\begin{document}


\title{\boldmath Measurement of the form factors of the decay $B^0\to
  D^{*-}\ell^+\nu_\ell$ and determination of the CKM matrix element
  $|V_{cb}|$}

\affiliation{Budker Institute of Nuclear Physics, Novosibirsk}
\affiliation{University of Cincinnati, Cincinnati, Ohio 45221}
\affiliation{The Graduate University for Advanced Studies, Hayama}
\affiliation{Hanyang University, Seoul}
\affiliation{University of Hawaii, Honolulu, Hawaii 96822}
\affiliation{High Energy Accelerator Research Organization (KEK), Tsukuba}
\affiliation{Institute of High Energy Physics, Chinese Academy of Sciences, Beijing}
\affiliation{Institute of High Energy Physics, Vienna}
\affiliation{Institute of High Energy Physics, Protvino}
\affiliation{Institute for Theoretical and Experimental Physics, Moscow}
\affiliation{J. Stefan Institute, Ljubljana}
\affiliation{Kanagawa University, Yokohama}
\affiliation{Institut f\"ur Experimentelle Kernphysik, Karlsruher Institut f\"ur Technologie, Karlsruhe}
\affiliation{Korea Institute of Science and Technology Information, Daejeon}
\affiliation{Korea University, Seoul}
\affiliation{Kyungpook National University, Taegu}
\affiliation{\'Ecole Polytechnique F\'ed\'erale de Lausanne (EPFL), Lausanne}
\affiliation{University of Maribor, Maribor}
\affiliation{Max-Planck-Institut f\"ur Physik, M\"unchen}
\affiliation{University of Melbourne, School of Physics, Victoria 3010}
\affiliation{Nagoya University, Nagoya}
\affiliation{Nara Women's University, Nara}
\affiliation{National Central University, Chung-li}
\affiliation{National United University, Miao Li}
\affiliation{Department of Physics, National Taiwan University, Taipei}
\affiliation{H. Niewodniczanski Institute of Nuclear Physics, Krakow}
\affiliation{Nippon Dental University, Niigata}
\affiliation{Niigata University, Niigata}
\affiliation{University of Nova Gorica, Nova Gorica}
\affiliation{Novosibirsk State University, Novosibirsk}
\affiliation{Osaka City University, Osaka}
\affiliation{Panjab University, Chandigarh}
\affiliation{Saga University, Saga}
\affiliation{University of Science and Technology of China, Hefei}
\affiliation{Seoul National University, Seoul}
\affiliation{Sungkyunkwan University, Suwon}
\affiliation{School of Physics, University of Sydney, NSW 2006}
\affiliation{Tata Institute of Fundamental Research, Mumbai}
\affiliation{Excellence Cluster Universe, Technische Universit\"at M\"unchen, Garching}
\affiliation{Tohoku Gakuin University, Tagajo}
\affiliation{Tohoku University, Sendai}
\affiliation{Department of Physics, University of Tokyo, Tokyo}
\affiliation{Tokyo Metropolitan University, Tokyo}
\affiliation{Tokyo University of Agriculture and Technology, Tokyo}
\affiliation{IPNAS, Virginia Polytechnic Institute and State University, Blacksburg, Virginia 24061}
\affiliation{Wayne State University, Detroit, Michigan 48202}
\affiliation{Yonsei University, Seoul}
  \author{W.~Dungel}\affiliation{Institute of High Energy Physics, Vienna} 
  \author{C.~Schwanda}\affiliation{Institute of High Energy Physics, Vienna} 
  \author{I.~Adachi}\affiliation{High Energy Accelerator Research Organization (KEK), Tsukuba} 
  \author{H.~Aihara}\affiliation{Department of Physics, University of Tokyo, Tokyo} 
  \author{T.~Aushev}\affiliation{\'Ecole Polytechnique F\'ed\'erale de Lausanne (EPFL), Lausanne}\affiliation{Institute for Theoretical and Experimental Physics, Moscow} 
  \author{T.~Aziz}\affiliation{Tata Institute of Fundamental Research, Mumbai} 
  \author{A.~M.~Bakich}\affiliation{School of Physics, University of Sydney, NSW 2006} 
  \author{V.~Balagura}\affiliation{Institute for Theoretical and Experimental Physics, Moscow} 
  \author{E.~Barberio}\affiliation{University of Melbourne, School of Physics, Victoria 3010} 
  \author{M.~Bischofberger}\affiliation{Nara Women's University, Nara} 
% ...
\collaboration{The Belle Collaboration}

\date{\today}

\begin{abstract}
  This article describes a determination of the Cabibbo-Kobayashi-Maskawa
matrix element~$|V_{cb}|$ from the decay $B^0\to
D^{*-}\ell^+\nu_\ell$ using 711 fb$^{-1}$ of Belle data collected near
the $\Upsilon(4S)$~resonance. We simultaneously measure the
product of the form factor normalization $\mathcal{F}(1)$ and the matrix 
element $|V_{cb}|$ as well as the three parameters $\rho^2$, $R_1(1)$ 
and $R_2(1)$, which determine the form factors of this decay in the 
framework of the Heavy Quark Effective Theory. The results, 
based on about 120,000 reconstructed $B^0\to D^{*-}\ell^+\nu_\ell$ decays, are
$\rho^2=1.214\pm 0.034\pm 0.009$, $R_1(1)=1.401\pm 0.034\pm 0.018$,
$R_2(1)=0.864\pm 0.024\pm 0.008$ and $\mathcal{F}(1)|V_{cb}|=(34.6\pm
0.2\pm 1.0)\times 10^{-3}$. The branching fraction of $B^0\to D^{*-}\ell^+\nu_\ell$
is measured at the same time; we obtain a value of $\mathcal{B}(B^0 \to 
D^{*-}\ell^+ \nu_\ell) = (4.58 \pm 0.03 \pm 0.26) \%$. The errors 
correspond to the statistical and systematic uncertainties. These results give
the most precise determination of the form factor parameters and 
$\mathcal{F}(1)|V_{cb}|$ to date. In addition, a direct, model-independent
determination of the form factor shapes has been carried out.

\end{abstract}

\pacs{}

\maketitle

\section{Introduction}

The study of the decay~$B^0\to D^{*-}\ell^+\nu_\ell$ is important for 
several reasons. The total rate is proportional to the magnitude of the
Cabibbo-Kobayashi-Maskawa(CKM) matrix element 
$V_{cb}$~\cite{Kobayashi:1973fv,Cabibbo:1963yz} squared.
Experimental investigation of the form factors of the decay can
check theoretical models and possibly provide input to more detailed 
theoretical approaches. In addition,
$B^0\to D^{*-}\ell^+\nu_\ell$ is a major background for charmless 
semileptonic $B$~decays, such as $B\to \pi \ell\nu$, or semileptonic
$B$~decays with large missing energy, including $B \to D^* \tau \nu$. 
Precise knowledge of the form factors in the~$B^0\to D^{*-}\ell^+\nu_\ell$ 
decay will thus help to reduce systematic uncertainties in these analyses.

This article is organized as follows: After introducing the
theoretical framework for the study of~$B^0\to D^{*-}\ell^+\nu_\ell$ decays 
in Section~\ref{sec:theo}, the experimental procedure is presented in detail in Section~
\ref{sec:expProcedure}. This is followed by a discussion of
our results and the systematic uncertainties assuming the form factor
parameterization of Caprini {\it et al.}~\cite{Caprini:1997mu} in Section
~\ref{sec:1dfit}. Finally, a measurement of the form factor shapes is 
described in Section~\ref{sec:shapes}.

This paper supersedes our previous result~\cite{Abe:2001cs}, based on a subset of the 
data used in this analysis.

\section{Theoretical framework}
\label{sec:theo}
\subsection{Kinematic variables} 

\begin{figure}[floatfix]
  \includegraphics[width=5cm]{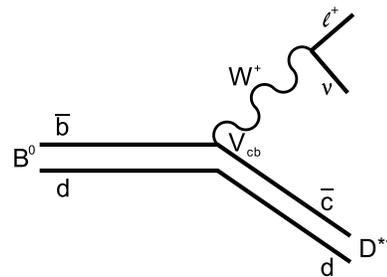}
  \caption{Quark-level Feynman diagram for the decay $B^0\to
    D^{*-}\ell^+\nu_\ell$.} \label{fig:1}
\end{figure}

The decay $B^0\to D^{*-}\ell^+\nu_\ell$~\cite{ref:0} proceeds
through the tree-level transition shown in Fig.~\ref{fig:1}. Below
we will follow the formulation proposed in reviews
~\cite{Neubert:1993mb,Richman:1995wm}, where the kinematics 
of this process are fully characterized by four variables as
discussed below.

The first is a function of the momenta of the $B$ and $D^{*}$ mesons,
labeled $w$ and defined by
\begin{equation}
  w=\frac{P_B\cdot P_{D^*}}{m_B
    m_{D^*}}=\frac{m_B^2+m_{D^*}^2-q^2}{2m_Bm_{D^*}}~,
    \label{eq:w}
\end{equation}
where $m_B$ and $m_{D^*}$ are the masses of the $B$ and the $D^*$
mesons (5.279 and 2.010 GeV/$c^2$, respectively~\cite{Amsler:2008zzb}), 
$P_B$ and $P_{D^*}$ are their four-momenta, and $q^2=(P_\ell+P_\nu)^2$. 
In the $B$~rest frame the expression for $w$
reduces to the Lorentz boost $\gamma_{D^*}=E_{D^*}/m_{D^*}$. The
ranges of $w$ and $q^2$ are restricted by the kinematics of the decay,
with $q^2_\mathrm{min} = 0$ corresponding to
\begin{equation}
  w_\mathrm{max}=\frac{m_B^2 + m_{D^*}^2}{2m_B m_{D^*}} \approx 1.504~,
\end{equation}
and $w_\mathrm{min}=1$ to
\begin{equation}
q_\mathrm{max}^2=(m_B-m_{D^*})^2 \approx 10.69~\mathrm{GeV}^2~.
\end{equation}
The point~$w=1$ is also referred to as zero recoil.

\begin{figure}[floatfix]
  \includegraphics[width=0.8\columnwidth]{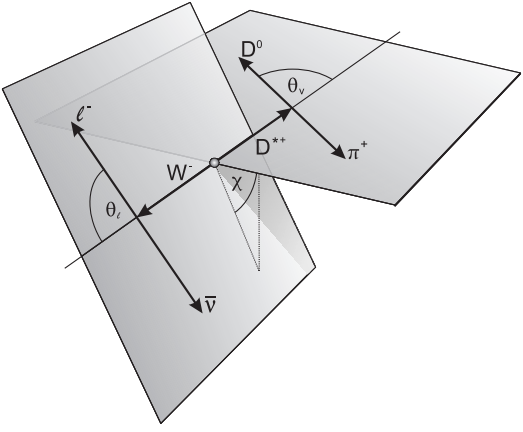}
  \caption{Definition of the angles $\theta_\ell$, $\theta_V$ and
    $\chi$ for the decay $B^0\to D^{*-}\ell^+\nu_\ell$, $D^{*-}\to\bar
    D^0\pi_s^-$.} \label{fig:2}
\end{figure}

The remaining three variables are the angles shown in Fig.~\ref{fig:2}:
\begin{itemize} 
\item{$\theta_\ell$, the angle between the direction of the lepton
  and the direction opposite the $B$ meson in the virtual $W$~rest frame;}
\item{$\theta_V$, the angle between the direction of the $D$~meson and
   the direction opposite the $B$ meson in the $D^*$ rest frame;}
\item{$\chi$, the angle between the plane formed by the $D^{*}$ decay and
  the plane formed by the $W$~decay, defined in the $B$ meson rest frame.}
\end{itemize} 

\subsection{Four-dimensional decay distribution}

Three helicity amplitudes, labeled $H_{+}$, $H_{-}$, and $H_{0}$,
can be used to describe the Lorentz structure of the $B^0\to D^{*-}\ell^+\nu_\ell$~decay
amplitude. These quantities correspond to the three
polarization states of the $D^*$, two transverse and one
longitudinal. When neglecting the lepton mass, {\it i.e.}, considering only electrons and muons, these
amplitudes are expressed in terms of the three functions $h_{A_1}(w)$,
$R_1(w)$, and $R_2(w)$ as follows~\cite{Neubert:1993mb}:
\begin{equation}
  H_i(w)=m_B\frac{R^*(1-r^2)(w+1)}{2\sqrt{1-2wr+r^2}}h_{A_1}(w)\tilde{H}_i(w)~,
\end{equation}
where
\begin{eqnarray}
  \tilde{H}_{\mp} & = &
  \frac{\sqrt{1-2wr+r^2}\left(1\pm\sqrt{\frac{w-1}{w+1}}
    R_1(w)\right)}{1-r}~, \\
  \tilde{H}_0 & = & 1+\frac{(w-1)(1-R_2(w))}{1-r}~,
\end{eqnarray}
with $R^*=(2\sqrt{m_B m_{D^*}})/(m_B+m_{D^*})$ and
$r=m_{D^*}/m_B$. The functions $R_1(w)$ and $R_2(w)$ are defined in
terms of the axial and vector form factors as,
\begin{equation}
  A_2(w)=\frac{R_2(w)}{R^{*2}}\frac{2}{w+1}A_1(w)~,
\end{equation}
\begin{equation}
  V(w)=\frac{R_1(w)}{R^{*2}}\frac{2}{w+1}A_1(w)~.
\end{equation}
By convention, the function $h_{A_1}(w)$ is defined as
\begin{equation}
  h_{A_1}(w)=\frac{1}{R^*}\frac {2}{w+1} A_1(w)~.
\end{equation}
The axial form factor $A_1(w)$ dominates for $w\to 1$. Furthermore, in the
limit of infinite $b$- and $c$-quark masses, a single form factor
describes the decay, the so-called Isgur-Wise
function~\cite{Isgur:1989vq,Isgur:1989ed}.

In terms of the three helicity amplitudes, the fully differential decay rate
is given by
\begin{equation}
  \begin{split}
    & \frac{d^4\Gamma(B^0\to
      D^{*-}\ell^+\nu_\ell)}{dw \,d(\cos\theta^{}_\ell)\,d(\cos\theta^{}_V)\,d\chi}=
      \frac{6m_Bm_{D^*}^2}{8(4\pi)^4}\sqrt{w^2-1}\\
    & (1-2wr+r^2)G_F^2|V_{cb}|^2 \big\{(1-\cos\theta_\ell)^2\sin^2\theta_VH^2_+(w)\\
    & +(1+\cos\theta_\ell)^2\sin^2\theta_VH^2_-(w)  +4\sin^2\theta_\ell\cos^2\theta_VH^2_0(w)\\
    & -2\sin^2\theta_\ell\sin^2\theta_V\cos 2\chi H_+(w)H_-(w) \\
    & -4\sin\theta_\ell(1-\cos\theta_\ell)\sin\theta_V\cos\theta_V\cos\chi  H_+(w)H_0(w) \\
    & +4\sin\theta_\ell(1+\cos\theta_\ell)\sin\theta_V\cos\theta_V\cos\chi  H_-(w)H_0(w)\big\}~, \label{eq:2_1}
  \end{split}
\end{equation}
with $G_F=(1.16637\pm 0.00001)\times 10^{-5}$~$\hbar\,c^2\, \mathrm{GeV}^{-2}$. 
Four one-dimensional decay distributions can be obtained by
integrating this decay rate over all but one of the four variables,
$w$, $\cos\theta_\ell$, $\cos\theta_V$, or $\chi$. The differential
decay rate as a function of $w$ is
\begin{equation}
  \frac{d\Gamma}{dw}=\frac{G^2_F}{48\pi^3}m^3_{D*}\big(m_B-m_{D^*}\big)^2\mathcal{G}(w)\mathcal{F}^2(w)|V_{cb}|^2~, \label{eq:2_2}
\end{equation}
where  
\begin{eqnarray*}
  &&\mathcal{F}^2(w)\mathcal{G}(w)=h_{A_1}^2(w)\sqrt{w^2-1}(w+1)^2
  \\
  &&\phantom{\mathcal{F}^2(w)}
  \times \left\{2\times\left[\frac{1-2wr+r^2}{(1-r)^2}\right]\left[1+R_1(w)^2\frac{w^2-1}{w+1}\right]\right.
  \\
  &&\phantom{\mathcal{F}^2(w)}
	\left. +\left[1+(1-R_2(w))\frac{w-1}{1-r}\right]^2\right\}~,
\end{eqnarray*}
and $\mathcal{G}(w)$ is a known phase space factor,
\begin{equation*}
  \mathcal{G}(w)=\sqrt{w^2-1}(w+1)^2\left[1+4\frac{w}{w+1}\frac{1-2wr+r^2}{(1-r)^2}\right].
\end{equation*}

A value of the form factor normalization $\mathcal{F}(1)=1$ is predicted
by Heavy Quark Symmetry (HQS)~\cite{Neubert:1993mb} in the infinite quark-mass limit. Lattice 
QCD can be utilized to calculate corrections to this limit. The most 
recent result obtained in unquenched lattice QCD is
$\mathcal{F}(1)=0.921\pm 0.013\pm 0.020$~\cite{Bernard:2008dn}.

\subsection{Form factor parameterization} \label{sub:parameterization}

A parameterization of form factors $h_{A_1}(w)$,
$R_1(w)$, and $R_2(w)$ can be obtained using
heavy quark effective theory (HQET). Perfect heavy quark symmetry
implies that $R_1(w)=R_2(w)=1$, {\it i.e.}, the form factors $A_2$ and
$V$ are identical for all values of $w$ and differ from $A_1$ only by
a simple kinematic factor. Corrections to this approximation have been
calculated in powers of $\Lambda_\mathrm{QCD}/m_b$ and the strong
coupling constant $\alpha_s$. Various parameterizations in powers of
$(w-1)$ have been proposed. We adopt the following expressions derived
by Caprini, Lellouch and Neubert~\cite{Caprini:1997mu},
\begin{eqnarray}
  h_{A_1}(w) & = &
  h_{A_1}(1)\big[1-8\rho^2z+(53\rho^2-15)z^2\nonumber\\
  && \phantom{ h_{A_1}(1)\big[}-(231\rho^2-91)z^3\big]~,  \label{eq:2_3} \\
  R_1(w) & = & R_1(1)-0.12(w-1)+0.05(w-1)^2, \label{eq:2_4} \\ 
  R_2(w) & = & R_2(1)+0.11(w-1)-0.06(w-1)^{2}, \label{eq:2_5}  
\end{eqnarray}
where $z=(\sqrt{w+1}-\sqrt{2})/(\sqrt{w+1}+\sqrt{2})$. In addition to the
form factor normalization $\mathcal{F}(1) = h_{A_1}(1)$, these expressions
contain three free parameters, $\rho^{2}$, $R_1(1)$, and $R_2(1)$. 
The values of these parameters cannot be calculated in a model-independent
manner. Instead, they have to be extracted by an analysis of experimental 
data.

\section{Experimental procedure}
\label{sec:expProcedure}
\subsection{Data sample and event selection}

The data used in this analysis were taken with the Belle
detector~\cite{unknown:2000cg} at the KEKB asymmetric-energy
$e^+e^-$~collider~\cite{Kurokawa:2001nw}. 
The Belle detector is a large-solid-angle magnetic
spectrometer that consists of a silicon vertex detector (SVD),
a 50-layer central drift chamber (CDC), an array of
aerogel threshold Cherenkov counters (ACC),
a barrel-like arrangement of time-of-flight
scintillation counters (TOF), and an electromagnetic calorimeter
comprised of CsI(Tl) crystals (ECL) located inside 
a superconducting solenoid coil that provides a 1.5~T
magnetic field.  An iron flux-return located outside of
the coil is instrumented to detect $K_L^0$ mesons and to identify
muons (KLM).  The detector is described in detail in 
Ref.~\cite{unknown:2000cg}.
Two inner detector configurations were used. A 2.0 cm beampipe
and a 3-layer silicon vertex detector were used for the first sample
of $152$ million $B\bar{B}$ pairs, while a 1.5 cm beampipe, a 4-layer
silicon detector and a small-cell inner drift chamber were used to record  
the remaining $620$ million $B\bar{B}$ pairs~\cite{svd2}.  

The data sample consists of 711~fb$^{-1}$ taken at the
$\Upsilon(4S)$~resonance, or about $772$ million $B\bar B$~events. 
Another 88~fb$^{-1}$ taken at 60~MeV below the resonance are
used to estimate the non-$B\bar B$ (continuum) background. The
off-resonance data is scaled by the integrated on- to off-resonance
luminosity ratio corrected for the $1/s$~dependence of the $e^+ e^- \to
q\bar{q}$~cross section.

This data sample contains events recorded with two different detector
setups as well as two different tracking algorithms and
large differences in the input files used for Monte Carlo generation. To
ensure that no systematic uncertainty appears due to inadequate consideration
of these differences, we separate the data sample into four distinct sets
labeled A (141~fb$^{-1}$), B (274~fb$^{-1}$), C (189~fb$^{-1}$) and D 
(107~fb$^{-1}$), where the number in parentheses indicates the integrated
luminosity corresponding to the individual samples.

Monte Carlo generated samples of $B\bar{B}$ decays equivalent to about three times the
integrated luminosity are used in this analysis. Monte Carlo simulated
events are generated with the {\tt Evtgen} program~\cite{Lange:2001uf}, and
full detector simulation based on {\tt GEANT}~\cite{Brun:1987ma} is
applied. QED final state radiation in $B\to X\ell\nu$~decays is added using
the {\tt PHOTOS} package~\cite{Barberio:1993qi}.

Hadronic events are selected based on the charged track multiplicity
and the visible energy in the calorimeter. The selection is described
in detail elsewhere~\cite{Abe:2001hj}. We also apply a
requirement on the ratio of the second to the zeroth Fox-Wolfram
moment~\cite{Fox:1978vu}, $R_2 < 0.4$, to reject continuum events.

\subsection{Event reconstruction}

Charged tracks are required to originate from the interaction point by
applying the following selections on the impact parameters in the $r-\phi$
and $z$ directions: $dr<2$~cm and $|dz|<4$~cm, respectively. In addition, we
demand at least one associated hit in the SVD detector. For pion and
kaon candidates, the Cherenkov light yield from the ACC, the
time-of-flight information from TOF, and $dE/dx$ from the CDC are required
to be consistent with the appropriate mass hypothesis.

Neutral $D$~meson candidates are reconstructed in the $D^0\to K^-\pi^+$
decay channel. We fit the charged tracks to a common vertex
and reject the $D^0$~candidate if the $\chi^2$-probability is below
$10^{-3}$. The reconstructed $D^0$~mass is required to lie within 
$\pm 13.75$~MeV/$c^2$ of the nominal $D^0$ mass of 1.865~GeV/$c^2$
~\cite{Amsler:2008zzb}, corresponding to about 2.5 times the experimental
resolution measured from data.

The $D^0$~candidate is combined with an additional charged pion
(oppositely charged with respect to the kaon candidate) to form a
$D^{*+}$~candidate. Due to the kinematics of the $D^{*+}$ decay, the 
momentum of this pion does not exceed 350 MeV/$c$. It is therefore referred 
to as the ``slow'' pion,~$\pi^+_s$. No impact parameter or SVD hit requirements are
applied for $\pi_s$. Again, a vertex fit is performed and the same
vertex requirement is applied. The invariant mass difference between the 
$D^*$ and the $D$~candidates, $\Delta m = m_{D^*} - m_{D^0}$, is 
required to be less than 165 MeV/$c^2$. This selection is tightened 
after the background estimation described below. Additional continuum 
suppression is achieved by requiring that the $D^*$~momentum in the c.m.\ frame
be below 2.45~GeV/$c$.

Finally, the $D^*$~candidate is combined with an oppositely charged
lepton (electron or muon). Electron candidates are identified using
the ratio of the energy detected in the ECL to the track momentum, the
ECL shower shape, position matching between track and ECL cluster, the
energy loss in the CDC, and the response of the ACC~counters. Muons are
identified based on their penetration range and transverse scattering
in the KLM~detector. In the momentum region relevant to this analysis,
charged leptons are identified with an efficiency of about 90\% while
the probability to misidentify a pion as an electron (muon) is 0.25\%
(1.4\%)~\cite{Hanagaki:2001fz,Abashian:2002bd}. Lepton tracks have to
be associated with at least one SVD hit. In the laboratory frame, the 
momentum of the electron (muon) is required to be greater than 0.30 GeV/$c$ 
(0.60 GeV/$c$). We also require the
lepton momentum in the c.m.\ frame to be less than 2.4~GeV/$c$ to 
reject continuum. More stringent lepton requirements are imposed later
in the analysis. 

For electron candidates we attempt bremsstrahlung
recovery by searching for photons within a cone of $3$ degrees around the electron track.
If such a photon is found, it is merged with the electron and the sum of the momenta
is taken to be the lepton momentum.

\subsection{Background estimation} \label{sec:3c}

\begin{figure*}[floatfix]
  \includegraphics[width=0.85\columnwidth]{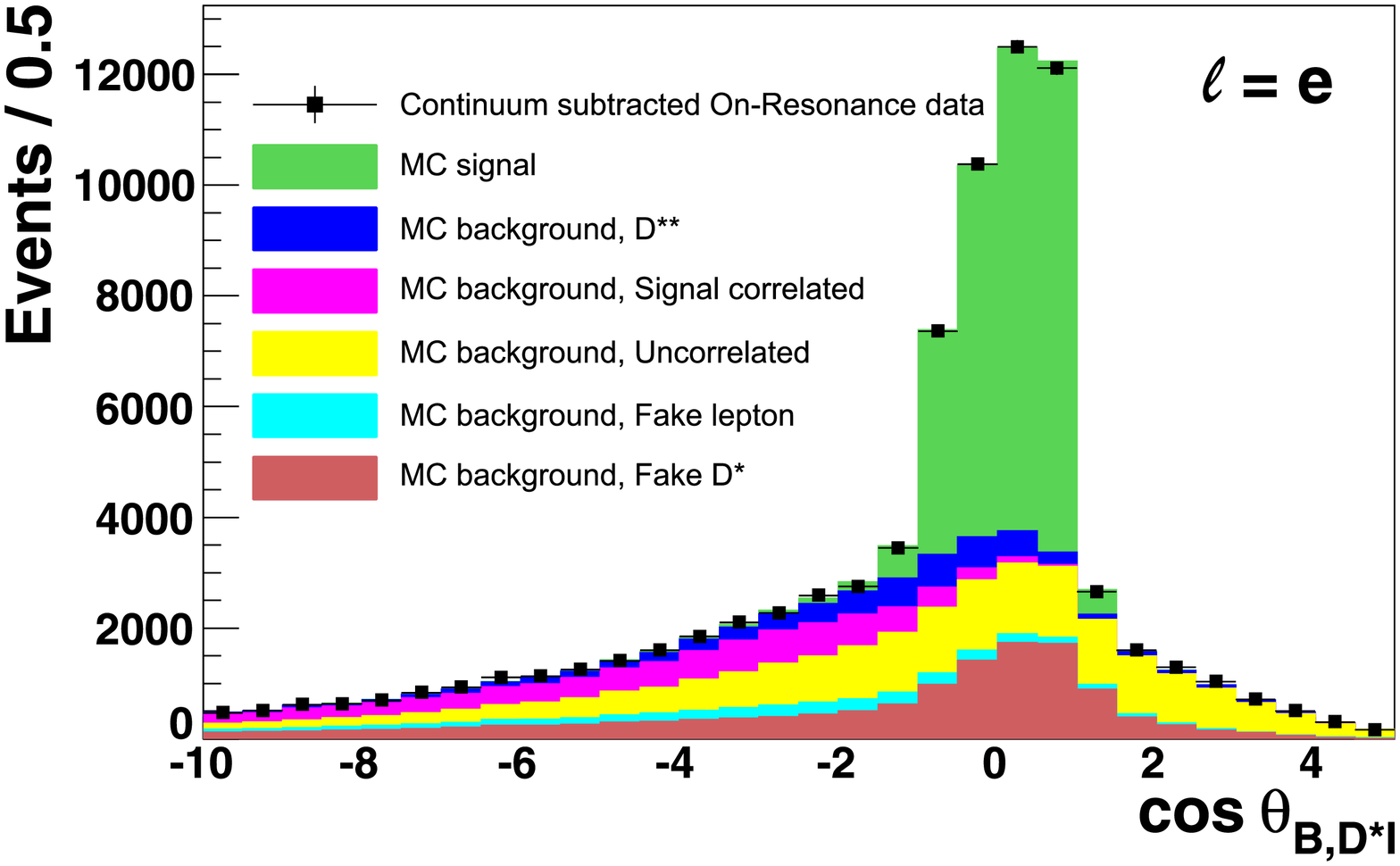}
  	\hspace{0.1\columnwidth}
  \includegraphics[width=0.85\columnwidth]{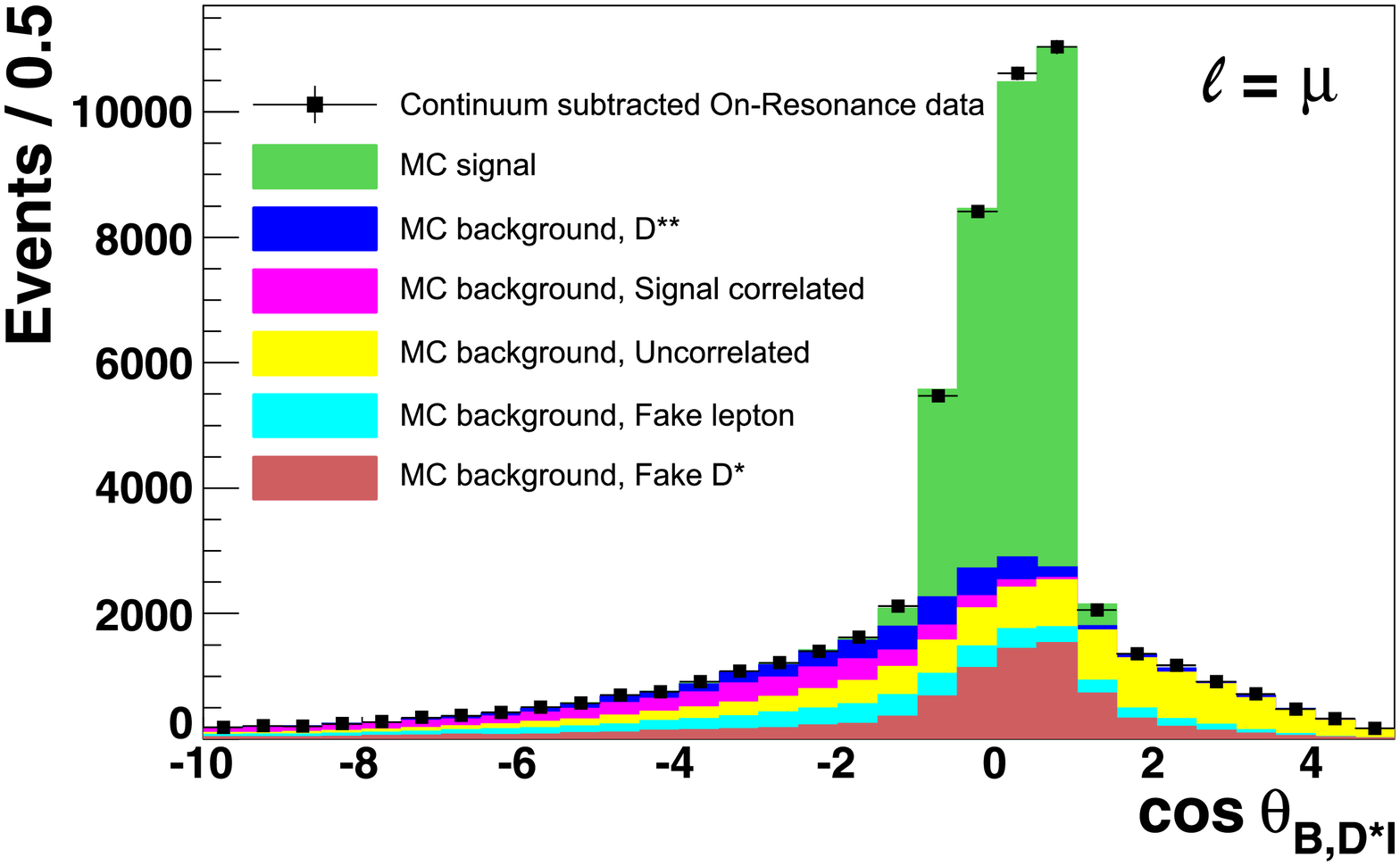}\\   
  \includegraphics[width=0.85\columnwidth]{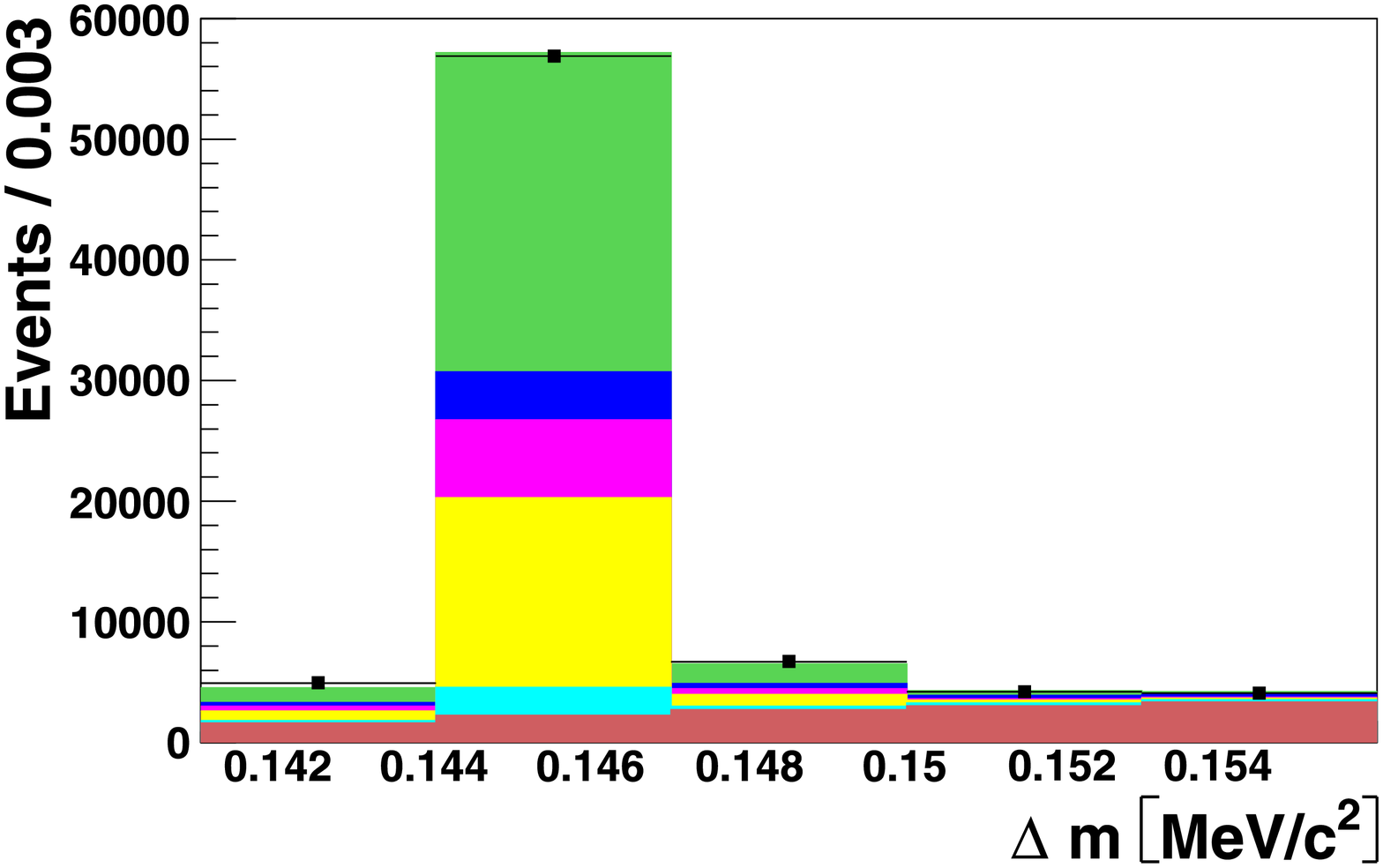}
  	\hspace{0.1\columnwidth}
  \includegraphics[width=0.85\columnwidth]{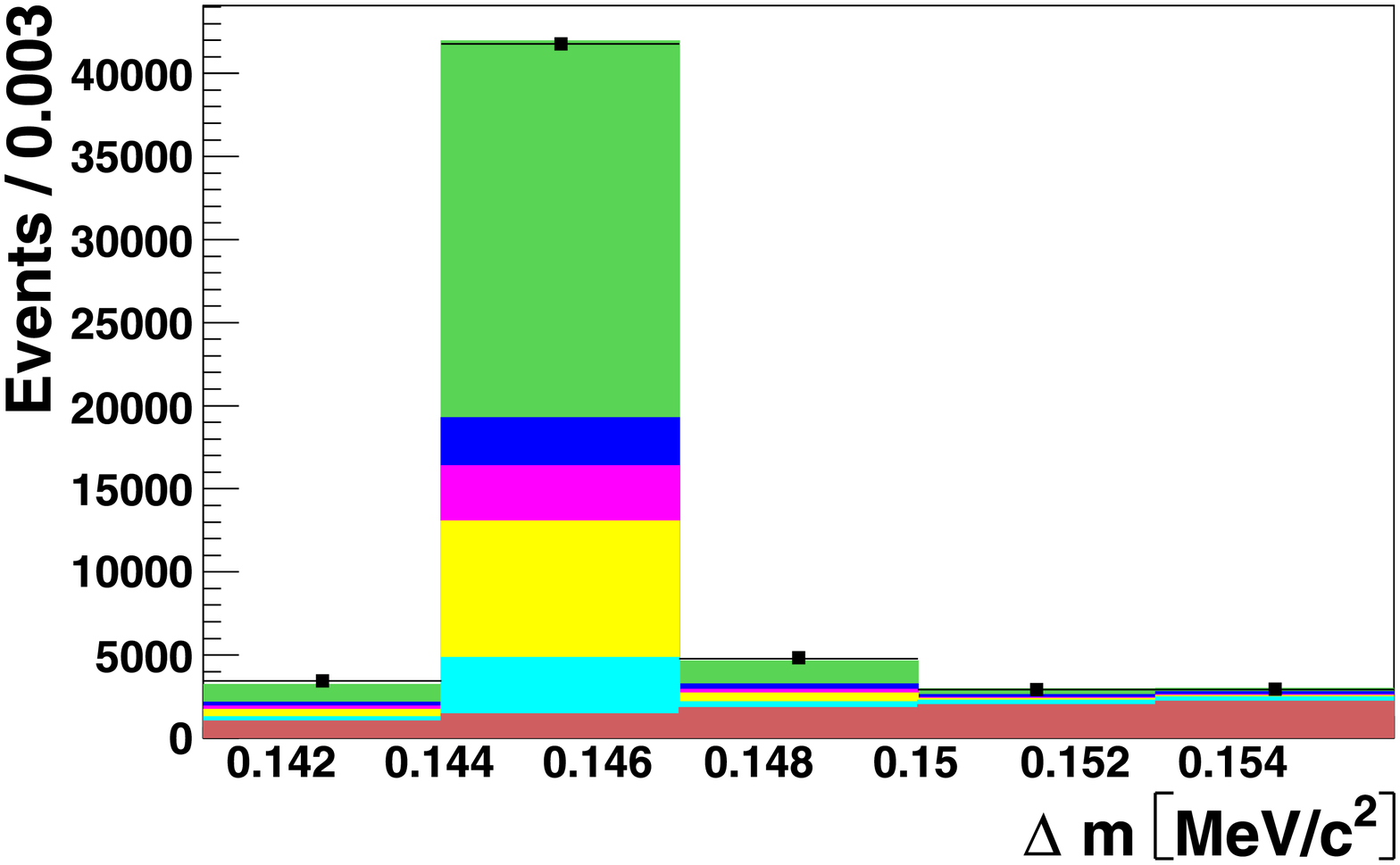}\\   
  \includegraphics[width=0.85\columnwidth]{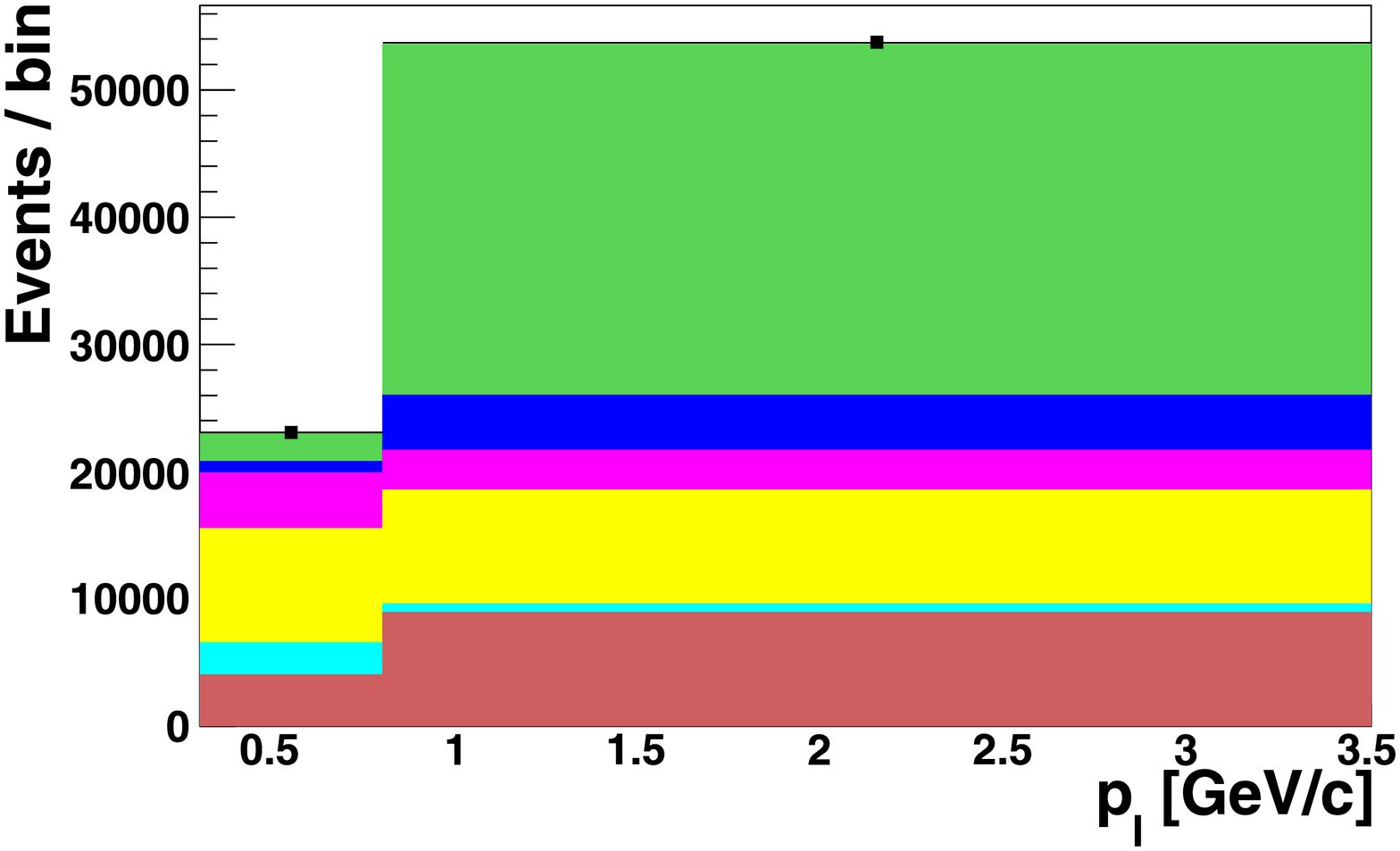}
  	\hspace{0.1\columnwidth}
  \includegraphics[width=0.85\columnwidth]{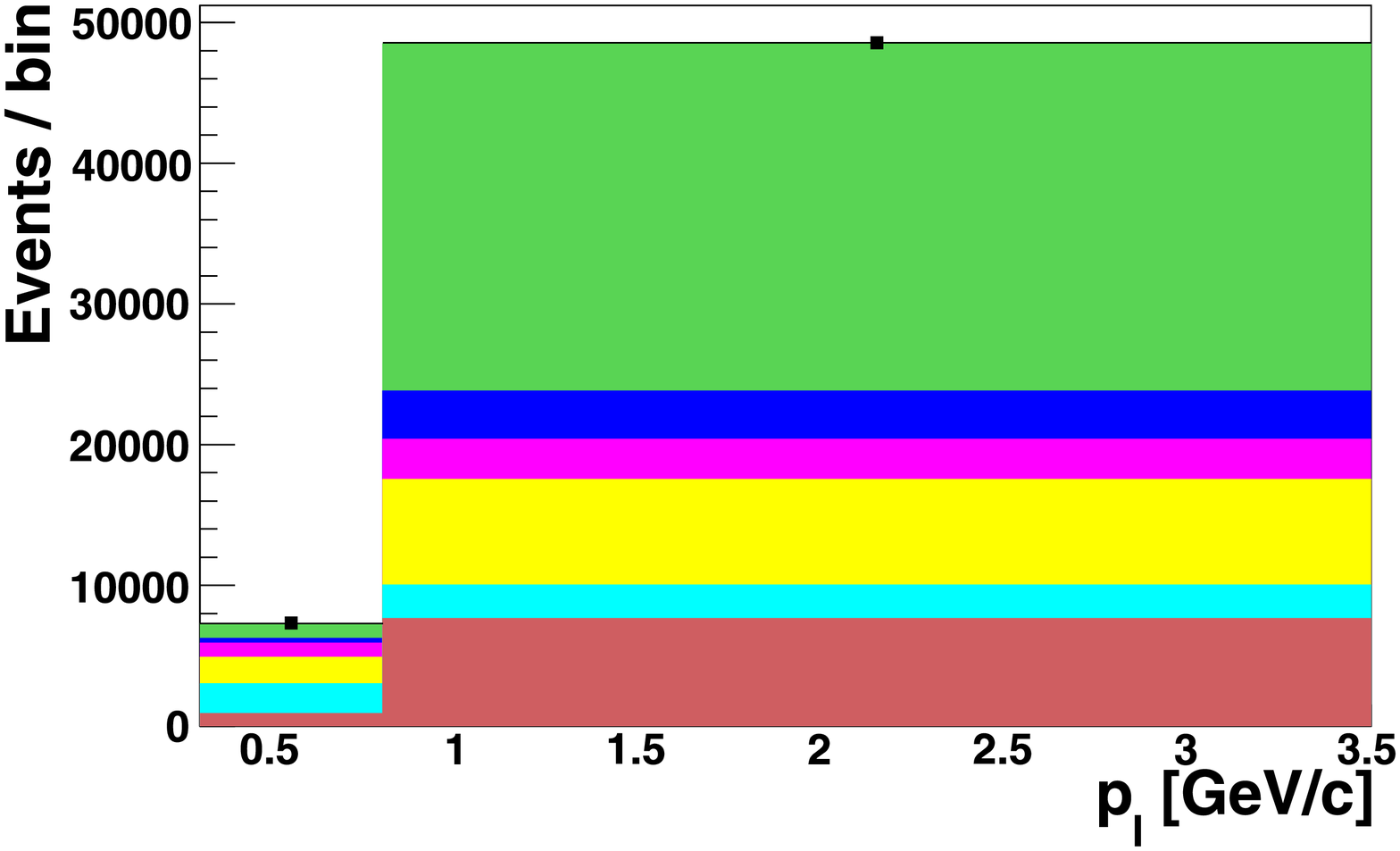} 
  \caption{Result of the fits to the ($\cos\theta_{B,D^{*0}\ell}$, 
    $\Delta m$, $p_\ell$)~distributions in the $e$ mode (left) and 
    $\mu$ mode (right) of the sub-sample B. The bin boundaries are
    discussed in the text. The points 
    with error bars are continuum-subtracted on-resonance data. Where 
    not shown, the uncertainties are smaller than the black markers.
    The histograms are, top to bottom, the signal component,
    $D^{**}$ background, signal correlated background, uncorrelated
    background, fake $\ell$ component and fake $D^*$ component.}
  \label{fig:TFFResultChan0}
\end{figure*}

Because we do not reconstruct the other $B$~meson in the event, the
$B$ momentum is \textit{a priori} unknown. However, in the c.m.\
frame, one can show that the $B$~direction lies on a cone around the
$(D^*\ell)$-axis~\cite{ref:2} with an opening angle $2\cos\theta_{B,D^*\ell}$
defined by:
\begin{equation}
  \cos\theta_{B,D^*\ell}=\frac{2E^{\,*}_BE^{\,*}_{D^*\ell}-m^2_B-m^2_{D^*\ell}}{2|\vec
  p^{\,*}_B||\vec p^{\,*}_{D^*\ell}|}~. \label{eq:3_1}
\end{equation}
In this expression, $E^{\,*}_B$ is half of the c.m.\ energy and $|\vec p^{\,*}_B|$ is
$\sqrt{E^{\,*2}_B-m^2_B}$. The quantities $E^{\,*}_{D^*\ell}$, $\vec
p^{\,*}_{D^*\ell}$ and $m_{D^*\ell}$ are calculated from the reconstructed
$D^*\ell$~system.

This cosine is also a powerful discriminator between signal and
background: Signal events should lie in the interval $(-1,1)$,
although -- due to finite detector resolution -- about 5\% of the signal is
reconstructed outside this interval. The background, on the other hand,
does not have this restriction. 

The signal lies predominantly in the region defined by 144 MeV/$c^2$ $< \Delta m <$ 147 MeV/$c^2$
and $p_\ell > 0.8$ GeV/$c$ ($p_\ell > 0.85$ GeV/$c$) for electrons (muons).
The region outside these thresholds can be used to estimate the background
level.

We therefore perform a fit to the three-dimensional 
$(\cos\theta_{B,D^*\ell}$,~$\Delta m$,~$p_\ell$)~distributions. The 
$\cos\theta_{B,D^{*}\ell}$ range between $-10$ and $5$ is divided into 30 bins.
The $\Delta m$ ($p_\ell$) range is divided into five (two) bins, with
bin boundaries at $141, 144, 147, 150, 153, 156$ MeV ($0.3, 0.8, 3.5$ GeV/$c$
for electrons and $0.6, 0.85, 3.5$ GeV/$c$ for muons).

The background contained in the final sample has the following six
components:
\begin{enumerate}

\item $D^{**}$: background from $B\to\bar D^{**}\ell^+\nu$~decays with
  $\bar D^{**}\to D^{*}\pi$ or $\bar D^{**}\to D\pi$ and from 
  non-resonant $B\to D^{*}\pi\ell^+\nu$ events, where the lepton has been
  correctly identified;

\item correlated background: background from processes other than
  $B\to\bar D^{**}\ell^+\nu$ decays in which the $D^*$ and the lepton originate 
  from the same $B$~meson, {\it e.g}, $B^0\to D^{*-}\tau^+\nu$, 
  $\tau^+\to\mu^+\nu\bar{\nu}$;

\item uncorrelated background: the $D^*$ and the lepton come from
  different $B$~mesons and the lepton is not from a 
  $B\to\bar D^{**}\ell^+\nu$~decay;

\item fake lepton: the charged lepton candidate is a misidentified hadron while
 the $D^*$~candidate may or may not be correctly reconstructed;

\item fake $D^*$: the $D^*$~candidate is misreconstructed; the lepton
  candidate is identified correctly, but it is not from a
  $B\to\bar D^{**}\ell^+\nu$~decay;

\item continuum: background from $e^+e^- \to q\bar{q}$ $(q=u,d,s,c)$ processes.

\end{enumerate}

To model the $D^{**}$ component, which consists of a total of four
resonant ($D_1$, $D_0^*$, $D_1'$, $D_2^*$) and one non-resonant
$D^{*} \pi \ell \nu$ mode for both neutral and charged $B$ decays,
we reweight the branching ratios of each subcomponent to match the values
reported by the Particle Data Group~\cite{Amsler:2008zzb}. For the 
resonant parts, only products of branching ratios $\mathcal{B}(B\to 
D^{**} \ell \nu)\times \mathcal{B}(D^{**} \to D^{(*)}\pi)$ are available
and consequently we reweight these products. The shape of the $D^{**}$
momentum distributions is also reweighted in 22 bins of $q^2$ to match the
predictions of the LLSW model~\cite{Leibovich:1997em,Urquijo:2006wd}. 

All of the background components are modeled by MC simulation except 
for continuum events; these are modeled by off-resonance data.
For muon events, the shape of the fake lepton background is corrected by the
ratio of the pion fake rate in the experimental data over the same
quantity in the Monte Carlo, as measured using
$K_S^0 \to\pi^+\pi^-$~decays. 
The lepton identification efficiency is corrected by the ratio between experimental
data and Monte Carlo in $2\gamma \to ee/\mu\mu$ events~\cite{Hanagaki:2001fz,Abashian:2002bd}. 
The ($\cos\theta_{B,D^*\ell}$, $\Delta m$, $p_\ell$)~distribution in the data is
fitted using the {\tt TFractionFitter} algorithm~\cite{Barlow:1993dm} within 
{\tt ROOT}~\cite{Brun:1996ro}. The fit is done separately in each of eight subsamples defined by the experiment 
range and the lepton type. The results are given in Table~\ref{tab:BackgroundFractions}.
Figure~\ref{fig:TFFResultChan0} shows plots of the projections in $\cos\theta_{B,D^*\ell}$ 
for subsample B.

In all fits, the continuum normalization is fixed to the on- to
off-resonance luminosity ratio, corrected for the $1/s$~dependence of
the $e^+e^-\to q\bar q$~cross section.
In general, the normalizations obtained by the fit agree well with the MC
expectations except for the $D^{**}$~component and the fake $\ell$ component, which
are overestimated in the MC. After the background determination only candidates satisfying
the requirements $-1 < \cos \theta_{B,D^*\ell} < 1$, $144~\mathrm{MeV}/c^2 < \Delta m < 
147~\mathrm{MeV}/c^2$ and $p_\ell>0.8$ GeV/$c$ ($p_\ell > 0.85$ GeV/$c$) for 
electrons (muons) are considered for further analysis.

\begin{table*}
  \begin{center}
    \renewcommand\arraystretch{1.2}
\begin{tabular}{l|@{\extracolsep{.2cm}}cccc}
\hline\hline
           & A, $K \pi, e$ & A, $K \pi, \mu$ & B, $K\pi, e$ & B, $K\pi, \mu$  \\
\hline
Num. Candidates &      14802 &            14203 &       29217 &      26894             \\

Signal events &     11609  $\pm$       181  &     11139  $\pm$       190  &     23029  $\pm$       280  &     21002  $\pm$       258  \\
\hline
    Signal ($\%$)&     78.43  $\pm$      1.22  &     78.43  $\pm$      1.34  &     78.82  $\pm$      0.96  &     78.09  $\pm$      0.96  \\

       $D^{**}$ ($\%$) &      5.63  $\pm$      0.78  &      4.02  $\pm$      0.86  &      4.32  $\pm$      0.66  &      3.90  $\pm$      0.60  \\

Signal correlated ($\%$) &      1.07  $\pm$      0.17  &      1.41  $\pm$      0.25  &      1.33  $\pm$      0.16  &      1.71  $\pm$      0.19  \\

Uncorrelated ($\%$) &      7.24  $\pm$      0.35  &      6.01  $\pm$      0.40  &      7.19  $\pm$      0.31  &      6.31  $\pm$      0.29  \\

    Fake $\ell$ ($\%$) &      0.36  $\pm$      0.17  &      1.99  $\pm$      0.34  &      0.50  $\pm$      0.17  &      2.10  $\pm$      0.23  \\

   Fake $D^*$ ($\%$) &      2.59  $\pm$      0.12  &      2.81  $\pm$      0.13  &      3.07  $\pm$      0.11  &      2.96  $\pm$      0.10  \\

 Continuum ($\%$) &      4.68  $\pm$      0.54  &      5.32  $\pm$      0.59  &      4.77  $\pm$      0.38  &      4.93  $\pm$      0.40  \\
\hline\hline
           & C, $K \pi, e$ & C, $K \pi, \mu$ & D, $K\pi, e$ & D, $K\pi, \mu$  \\
\hline
Num. Candidates &      22056 &      20428 &      15871 &      14719             \\

Signal events &     17301  $\pm$       240  &     15513  $\pm$       235  &     12365  $\pm$       189  &     11469  $\pm$       205  \\
\hline
    Signal ($\%$) &     78.44  $\pm$      1.09  &     75.94  $\pm$      1.15  &     77.91  $\pm$      1.19  &     77.92  $\pm$      1.39  \\

       $D^{**}$ ($\%$) &      5.15  $\pm$      0.71  &      5.22  $\pm$      0.71  &      4.54  $\pm$      0.72  &      4.67  $\pm$      0.86  \\

Signal correlated ($\%$) &      1.56  $\pm$      0.27  &      2.07  $\pm$      0.37  &      2.01  $\pm$      0.26  &      2.73  $\pm$      0.43  \\

Uncorrelated ($\%$) &      6.35  $\pm$      0.35  &      6.01  $\pm$      0.33  &      7.33  $\pm$      0.38  &      6.30  $\pm$      0.40  \\

    Fake $\ell$ ($\%$) &      0.75  $\pm$      0.18  &      2.26  $\pm$      0.28  &      0.30  $\pm$      0.19  &      1.68  $\pm$      0.38  \\

   Fake $D^*$ ($\%$) &      2.86  $\pm$      0.12  &      2.69  $\pm$      0.11  &      2.89  $\pm$      0.13  &      2.80  $\pm$      0.14  \\

 Continuum ($\%$) &      4.88  $\pm$      0.45  &      5.81  $\pm$      0.51  &      5.02  $\pm$      0.53  &      3.89  $\pm$      0.49  \\
\hline\hline

\end{tabular}  

  \end{center}
  \caption{The signal yield and the signal and background fractions (given in $\%$) 
  	for selected events passing the requirements $|\cos\theta_{B,D^*\ell}|<1$,
    $144~\mathrm{MeV}/c^2 < \Delta m < 147~\mathrm{MeV}/c^2$ and $p_\ell > 0.8$ GeV/$c$
    ($p_\ell > 0.85$ GeV/$c$) for electron (muon) channels.} 
  \label{tab:BackgroundFractions}
\end{table*}

\subsection{Kinematic variables}

\begin{figure}[floatfix]
  \includegraphics[width=0.8\columnwidth]{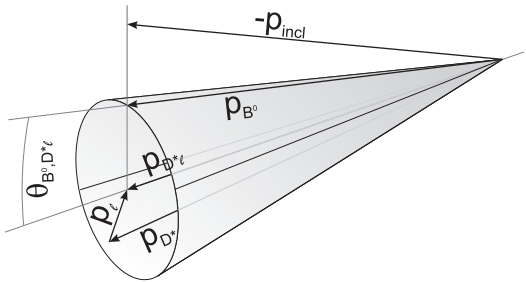}
  \caption{Reconstruction of the $B^0$~direction. Refer to the text
    for details.} \label{fig:6}
\end{figure}

To calculate the four kinematic variables defined in Eq.~\ref{eq:w} and Fig.~\ref{fig:2} 
-- $w$, $\cos\theta_\ell$, $\cos\theta_V$ and $\chi$ -- which characterize
the $B^0\to D^{*-}l^+\nu$~decay, we need to
determine the $B^0$~rest frame. The $B$~direction is already known to
lie on a cone around the $(D^*\ell)$-axis with an opening angle
$2\theta_{B,D^*\ell}$ in the c.m.\ frame, Equation~(\ref{eq:3_1}). To initially
determine the $B$~direction, we estimate the c.m.\ frame momentum
vector of the non-signal $B$ meson
by summing the momenta of the remaining particles in the
event ($\vec p^{\,*}_\mathrm{incl}$~\cite{ref:2}) and choose the
direction on the cone that minimizes the difference to $-\vec
p^{\,*}_\mathrm{incl}$, Fig.~\ref{fig:6}.

To obtain $\vec p^*_\mathrm{incl}$, we exclude tracks
passing far from the interaction point. The minimal requirements
depend on the transverse momentum of the track, $p_T$, and are set to
$dr>20$~cm ($15$~cm, $10$~cm) or $|dz|>100$~cm (50~cm, 20~cm) 
for a track $p_{T} < 250$~MeV/$c$ ($p_{T} < 500$~MeV/$c$, $p_{T} \ge 500$~MeV/$c$).
Track candidates that are compatible with a multiply reconstructed track generated by a 
low-momentum particle spiraling in the central drift chamber are also checked for
and only one of the multiple tracks is considered. 
Unmatched clusters  in the barrel region must have an energy greater than 50~MeV. 
For clusters in the forward (backward) region, the threshold is at 100~MeV
(150~MeV). We then compute $\vec p_\mathrm{incl}$ (in the laboratory
frame) by summing the 3-momenta of the selected particles,
\begin{equation}
  \vec p_\mathrm{incl}= \sum_i \vec p_i~,
\end{equation}
where the index $i$ stands for all particles passing the conditions above,
and transform this vector into the c.m.\ frame. Note that we do not introduce
any mass assumption for the charged particles. The energy component of
$p_\mathrm{incl}$ is determined by requiring $E^*_\mathrm{incl}$ to be 
$E^*_\mathrm{beam}=\sqrt{s}/2$.

With the $B^0$~rest frame reconstructed in this way, the resolutions
in the kinematic variables are found to be about 0.025, 0.049, 0.050
and 13.5$^\circ$ for $w$, $\cos\theta_\ell$, $\cos\theta_V$ and
$\chi$, respectively.

\section{Analysis based on the parameterization of Caprini {\it et al.}}
\label{sec:1dfit}
\subsection{Fit procedure}\label{section:exp-parametrizedFit}

Our main goal is to extract the following quantities: the product of the 
form factor normalization and $|V_{cb}|$,
$\mathcal{F}(1)|V_{cb}|$ (Eq.~\ref{eq:2_2}), and the three parameters
$\rho^2$, $R_1(1)$ and $R_2(1)$ that parameterize the form factors in
the HQET framework (Eqs.~\ref{eq:2_3}--\ref{eq:2_5}). For this,
we perform a binned $\chi^{2}$~fit to the $w$, $\cos\theta_\ell$, $\cos\theta_V$
and $\chi$~distributions over nearly the entire phase space. 
Instead of an unbinned fit, we fit the one-dimensional projections of $w$,
$\cos\theta_\ell$, $\cos\theta_V$ and $\chi$. This avoids the difficulty
of parameterizing the six background components and their correlations in
four dimensions. In addition, the one-dimensional projections have sufficient
statistics in each bin. However, this approach introduces bin-to-bin 
correlations that must be accounted for.

The distributions in $w$, $\cos\theta_\ell$, $\cos\theta_V$ and $\chi$
are divided into ten bins of equal width. The kinematically allowed
values of $w$ are between $1$ and $1.504$, but we restrict the
fit range to values between $1$ and $1.5$. In each sub-sample, there
are thus 40~bins to be used in the fit. In the following, we label
these bins with a single index $i$, $i=1,\dots,40$. The bins $i=1,\dots,10$
correspond to the bins of the $w$~distribution, $i=11,\dots,20$ to
$\cos\theta_\ell$, $i=21,\dots,30$ to $\cos\theta_V$, and $i=31,\dots,40$ to
the $\chi$ distribution.

The number of produced events in the bin~$i$, $N^\mathrm{th}_i$, is 
given by
\begin{equation}
  N^\mathrm{th}_i=N_{B^0}\mathcal{B}(D^{*+}\to
  D^0\pi^+)\mathcal{B}(D^0\to K^-\pi^+)\tau_{B^0}\Gamma_i~,
\end{equation}
where $N_{B^0}$ is the number of $B^0$~mesons in the data sample,
$\mathcal{B}(D^{*+}\to D^0\pi^+)$ and $\mathcal{B}(D^0\to K^-\pi^+)$ 
are the $D^*$ and $D$ branching ratios into the
final state under consideration~\cite{Amsler:2008zzb}, $\tau_{B^0}$ is the
$B^0$~lifetime~\cite{Amsler:2008zzb}, and $\Gamma_i$ is the width obtained
by integrating Eq.~\ref{eq:2_1} in the kinematic variable
corresponding to $i$ from the lower to the upper bin boundary (the
other kinematic variables are integrated over their full range). This
integration is numerical in the case of $w$ and analytic for the other
variables. The expected number of events $N^\mathrm{exp}_i$ is related
to $N^\mathrm{th}_i$ as follows
\begin{equation}
  N^\mathrm{exp}_i=\sum_{j=1}^{40}\left(R_{ij}\epsilon_jN^\mathrm{th}_j\right)+N^\mathrm{bkgrd}_i~.
  \label{eq:ApplicationOfEfficiencies}
\end{equation}
Here, $\epsilon_j$ is the probability that an event generated in the
bin~$j$ is reconstructed and passes all analysis cuts, and $R_{ij}$ is the
detector response matrix, \textit{i.e.}, it gives the probability that
an event generated in the bin~$j$ is observed in the bin~$i$. Both
quantities are calculated using MC~simulation. $N^\mathrm{bkgrd}_i$ is
the number of expected background events, estimated as described in
Sect.~\ref{sec:3c}.

Next, we calculate the variance~$\sigma^2_i$ of $N^\mathrm{exp}_i$. We
consider the following contributions: the Poissonian uncertainty in
$N^\mathrm{th}_i$; fluctuations related to the efficiency,
estimated by a binomial distribution with $N$ repetitions and known 
success probability $\epsilon_i$; a similar contribution
related to $R_{ij}$ using a multinomial distribution; and the uncertainty in the
background contribution $N^\mathrm{bkgrd}_i$. This yields the
following expression for $\sigma^2_i$,
\begin{eqnarray}
    \sigma^2_i &=&
    \sum_{j=1}^{40}
    \left[R^2_{ij}\epsilon^2_jN^\mathrm{th}_j
    +R^2_{ij}\frac{\epsilon_j(1-\epsilon_j)}{N_\mathrm{data}}(N^\mathrm{th}_j)^2+\right.\nonumber\\
    &&\frac{R_{ij}(1-R_{ij})}{N'_\mathrm{data}}\epsilon^2_j(N^\mathrm{th}_j)^2
    +R^2_{ij}\frac{\epsilon_j(1-\epsilon_j)}{N_\mathrm{MC}}(N^\mathrm{th}_j)^2+\nonumber\\
    &&\left. \frac{R_{ij}(1-R_{ij})}{N'_\mathrm{MC}}\epsilon^2_j(N^\mathrm{th}_j)^2\right]+\sigma^2(N^\mathrm{bkgrd}_i)~.
      \label{eq:sigmaErrorPropagation}
\end{eqnarray}
The first term is the Poissonian uncertainty in $N^\mathrm{th}_i$. The
second and third terms are the binomial and multinomial uncertainties,
respectively, related to the finite real data size, where $N_\mathrm{data}$
($N'_\mathrm{data}$) is the total number of decays (the number of
reconstructed decays) into the final state under consideration ($K\pi$, 
$e$ or $\mu$) in the real data. The quantities $\epsilon_i$
and $R_{ij}$ are calculated from a finite signal MC sample
($N_\mathrm{MC}$ and $N'_\mathrm{MC}$); the corresponding
uncertainties are estimated by the fourth and fifth terms. Finally,
the last term is the background contribution
$\sigma^2(N^\mathrm{bkgrd}_i)$, calculated as the sum of the different
background component variances. For each background component
defined in Section~\ref{sec:3c} we estimate its contribution by linear error
propagation of the results determined in Section~\ref{sec:3c}.
For continuum, we estimate the error in the on-resonance to off-resonance 
luminosity ratio to be 1.0\%. These variances give the diagonal elements of the
covariance matrix~$C_{ij}$.

In each sub-sample we calculate the off-diagonal elements of the
covariance matrix~$C_{ij}, i\ne j$ as $Np_{ij}-Np_ip_j$, where $p_{ij}$ is
the relative abundance of bin~$(i,j)$ in the 2-dimensional histogram
obtained by plotting the kinematic variables against each other,
$p_i$ is the relative number of entries in the 1-dimensional
distribution, and $N$ is the size of the sample. Covariances are
calculated for the signal and the different background components in
the MC samples, and added with appropriate normalizations.

The covariance matrix is inverted numerically within
ROOT~\cite{Brun:1996ro} and, labeling the electron and muon mode in
each sub-sample with the index~$k$, $\chi^2$~functions are calculated,
\begin{equation}
  \chi^2_k=\sum_{i,j}(N^\mathrm{obs}_{k,i}-
  N^\mathrm{exp}_{k,i})C^{-1}_{ij}(N^\mathrm{obs}_{k,j}-N^\mathrm{exp}_{k,j})~,
\end{equation}
where $N^\mathrm{obs}_{k,i}$ is the number of events observed in bin~$i$
in data sample $k$. We sum these two functions in each sub-sample 
and minimize the global $\chi^2$ with MINUIT~\cite{James:1975dr}.

We have tested this fit procedure using generic MC data
samples. All results are consistent with expectations and show no
indication of bias.

\subsection{Investigation of the efficiency of low momentum tracks}
\label{section:softPiCorrections}

The tracking efficiency of the Belle experiment is reproduced well by
MC simulations for tracks with momenta above 200~MeV/$c$, which we refer
to as ``high momentum tracks''. However, a significant portion of the 
momentum spectrum of the 
slow pions emitted in the $D^*$ decay lies below this boundary. For low
momenta, the effects of interactions with the detector material such as 
multiple scattering and energy loss become important and
might lead to a deviation between data and MC in the reconstruction
efficiency.

We use one half of the reconstructed $B \to D^* \ell \nu$ sample
to obtain corrections to the MC reconstruction efficiency in the low
momentum range, measured using real data. The second half is used
to perform the analysis with a statistically independent sample. 
The results of the background estimation shown
in Table~\ref{tab:BackgroundFractions} are those obtained in the 
samples used for the analysis. Both of the samples contain about 
120,000 signal events.

The sample used to investigate the efficiency of low momentum tracks
is divided into a total of six bins in $p_{\pi_s}$. The bin borders of
the first five are 50 MeV$/c$, 100 MeV$/c$, 125 MeV$/c$, 150 MeV$/c$,
175 MeV$/c$ and 200 MeV$/c$. The region beyond 200 MeV/$c$ defines the
sixth bin. By subtracting the background, we obtain
an estimate of the signal in data and form the ratio with the signal
in MC in each bin, $f_i = N_i^{\mathrm{data}} / N_i^{\mathrm{MC}}$.

The high momentum range is used as normalization, no efficiency correction
is applied there. In the lower momentum bins we obtain the ratios $\rho_{\pi_s, i}
= f_i / f_{\mathrm{max}}$, which are identical to the ratio of reconstruction
efficiencies in the bins $i$ and the high momentum region, $\rho_{\pi_s, i}
= \epsilon_i / \epsilon_{\mathrm{max}}$. We calculate this set of ratios for
the electron and muon modes and form the weighted average, separately for 
each of the four sub-samples. These values are applied as weights 
when filling the  MC histograms to correct the reconstruction efficiency.

Most systematic uncertainties cancel out in the ratios $\rho_{\pi_s,i}$.
Only the uncertainties in the various background components give a
small systematic contribution to the uncertainty.

This procedure assumes that the distribution of events in the $p_{\pi_s}$
spectrum is identical for data and MC. However, one of the aims of the 
analysis is to measure the form factor parameters that govern this
distribution. Therefore, an iterative procedure is adopted: we
calculate one set of corrections, apply them and perform the analysis
to determine $\mathcal{F}(1)|V_{cb}|$ and the form factor parameters. We
then calculate a new set of corrections using these results and repeat 
the analysis. The changes of the parameters during this iterative procedure
are small and vanish after the third iteration. We assign an additional systematic
uncertainty to our results based on the stability of the corrections
against changes in the form factor parameters. As will be shown in
Table~\ref{tab:SystematicBreakdown}, this is a negligibly small contribution.

\begin{table*}
	\begin{center}
\begin{tabular}{l|@{\extracolsep{.2cm}}ccc}
\hline\hline
 Sample A          &  $D^0 \to K \pi, \ell = e$             &  $D^0 \to K \pi, \ell = \mu$ &  total sample\\
\hline 
     $\rho^2$ &     1.248  $\pm$     0.102  $\pm$     0.022  &     1.285  $\pm$     0.114  $\pm$     0.028  &     1.259  $\pm$     0.076  $\pm$     0.019  \\

        $R_1(1)$ &     1.317  $\pm$     0.099  $\pm$     0.041  &     1.577  $\pm$     0.131  $\pm$     0.036  &     1.436  $\pm$     0.078  $\pm$     0.030  \\

        $R_2(1)$ &     0.804  $\pm$     0.076  $\pm$     0.017  &     0.768  $\pm$     0.093  $\pm$     0.020  &     0.795  $\pm$     0.058  $\pm$     0.015  \\

$ \mathcal{F}(1) |V_{cb} | \times 10^3$ &      34.8  $\pm$       0.5  $\pm$       1.2  &      34.6  $\pm$       0.6  $\pm$       1.2  &      34.7  $\pm$       0.4  $\pm$       1.2  \\

\hline
$\chi^2$ / n.d.f &      32.2  /      36.0  &            31.6  /      36.0  &      70.9  /      76.0            \\

$P_{\chi^2}$ &     0.651  &     0.676  &     0.643              \\

\hline\hline
 Sample B         &  $D^0 \to K \pi, \ell = e$             &   $D^0 \to K \pi, \ell = \mu$ &  total sample\\
\hline 
     $\rho^2$ &     1.169  $\pm$     0.079  $\pm$     0.011  &     1.167  $\pm$     0.088  $\pm$     0.016  &     1.168  $\pm$     0.059  $\pm$     0.011  \\

        $R_1(1)$ &     1.411  $\pm$     0.079  $\pm$     0.026  &     1.449  $\pm$     0.090  $\pm$     0.028  &     1.427  $\pm$     0.059  $\pm$     0.022  \\

        $R_2(1)$ &     0.902  $\pm$     0.054  $\pm$     0.011  &     0.859  $\pm$     0.061  $\pm$     0.013  &     0.882  $\pm$     0.041  $\pm$     0.010  \\

$ \mathcal{F}(1) |V_{cb} | \times 10^3$ &      34.4  $\pm$       0.4  $\pm$       1.1  &      33.9  $\pm$       0.4  $\pm$       1.1  &      34.2  $\pm$       0.3  $\pm$       1.1  \\
\hline
$\chi^2$ / n.d.f &      22.7  /      36.0  &      36.5  /      36.0  &       60.7  /      76.0           \\

$P_{\chi^2}$ &     0.958  &     0.443  &     0.900         \\

\hline\hline
   Sample C       & $D^0 \to K \pi, \ell = e$             &  $D^0 \to K \pi, \ell = \mu$ & total sample\\
\hline 
     $\rho^2$ &     1.226  $\pm$     0.088  $\pm$     0.011  &     1.262  $\pm$     0.101  $\pm$     0.016  &     1.239  $\pm$     0.066  $\pm$     0.011  \\

        $R_1(1)$ &     1.363  $\pm$     0.086  $\pm$     0.026  &     1.480  $\pm$     0.107  $\pm$     0.033  &     1.411  $\pm$     0.066  $\pm$     0.023  \\

        $R_2(1)$ &     0.891  $\pm$     0.062  $\pm$     0.012  &     0.851  $\pm$     0.076  $\pm$     0.015  &     0.876  $\pm$     0.048  $\pm$     0.012  \\

$ \mathcal{F}(1) |V_{cb} | \times 10^3$ &      34.4  $\pm$       0.5  $\pm$       1.1  &      33.9  $\pm$       0.5  $\pm$       1.1  &      34.2  $\pm$       0.3  $\pm$       1.1  \\

\hline
$\chi^2$ / n.d.f &      38.6  /      36.0  &      38.2  /      36.0  &      81.4  /      76.0           \\

$P_{\chi^2}$ &     0.352  &     0.370  &     0.314              \\
\hline\hline
Sample D          & $D^0 \to K \pi, \ell = e$             &  $D^0 \to K \pi, \ell = \mu$ & total sample\\
\hline 
     $\rho^2$ &     1.321  $\pm$     0.102  $\pm$     0.019  &     1.174  $\pm$     0.106  $\pm$     0.020  &     1.247  $\pm$     0.073  $\pm$     0.014  \\

        $R_1(1)$ &     1.448  $\pm$     0.109  $\pm$     0.041  &     1.230  $\pm$     0.089  $\pm$     0.031  &     1.330  $\pm$     0.069  $\pm$     0.027  \\

        $R_2(1)$ &     0.791  $\pm$     0.081  $\pm$     0.019  &     0.931  $\pm$     0.071  $\pm$     0.015  &     0.864  $\pm$     0.053  $\pm$     0.014  \\

$ \mathcal{F}(1) |V_{cb} | \times 10^3$ &      35.4  $\pm$       0.6  $\pm$       1.2  &      35.7  $\pm$       0.6  $\pm$       1.2  &      35.6  $\pm$       0.4  $\pm$       1.2  \\

\hline
$\chi^2$ / n.d.f &      25.1  /      36.0  &            42.0  /      36.0  &      70.1  /      76.0            \\

$P_{\chi^2}$ &     0.913  &     0.226  &     0.669              \\

\hline\hline
\end{tabular}

  \end{center}
  \caption{The fit results for the four sub-samples. The first two columns show results
  	obtained by investigating only the $e$ or the $\mu$ channel, the third column
  	is obtained by minimizing the sum of the $\chi^2$ values calculated for each channel.
  	Where given, the first error is statistical, and the second is systematic.} 
  \label{tab:ParResults1}
\end{table*}
\begin{figure*}[floatfix]
  \includegraphics[width=1.6\columnwidth]{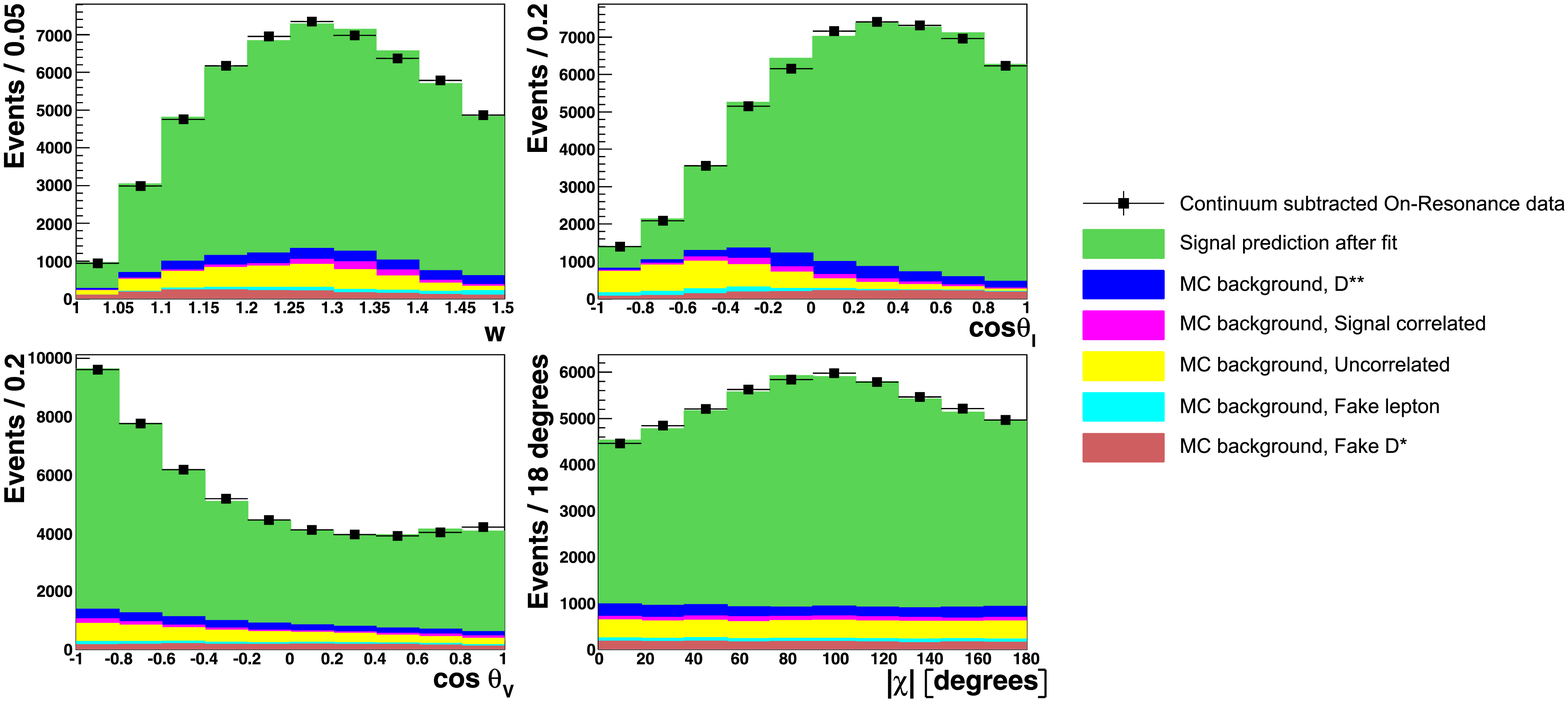}
  \caption{Result of the fit of the four kinematic variables in
    the sub-sample B. The electron and muon modes are added in this plot.
    The points with error bars are continuum subtracted on-resonance
    data. Where not shown, the uncertainties are smaller than the black markers.
    The histograms are, top to bottom, the signal component,
    $D^{**}$ background, signal correlated background, uncorrelated
    background, fake $\ell$ component and fake $D^*$ component.}
  \label{fig:fitResult}
\end{figure*}

\subsection{Results of the fits and investigation of the systematic uncertainties in the subsamples}

After applying all analysis cuts and subtracting backgrounds,
a total of $123,427\pm 636$ signal events are used for the analysis, 
divided into a total of four experimental sub-samples as mentioned above. 
The result of the fit to these data is shown in Fig.~\ref{fig:fitResult}
and Table~\ref{tab:ParResults1}. The $\chi^2$ per degree of freedom,
$\chi^2/\mathrm{n.d.f}$, of all fits is good. Table~\ref{tab:ParResults1}
also gives the $\chi^2$ probabilities or P-values, $P_{\chi^2}$.

To estimate the systematic uncertainties in these results,
we consider contributions from the following sources: uncertainties 
in the background component normalizations, uncertainty 
in the MC tracking efficiency, errors in the world average of $\mathcal{B}(D^{*+}\to D^0\pi^+)$
and $\mathcal{B}(D^0\to K^-\pi^+)$ as well as in the $\mathcal{B}(B\to D^{**}\ell\nu)$
components~\cite{Amsler:2008zzb}, uncertainties in the shape of the $w$ distribution of
$B\to D^{**}\ell\nu$ events based on the LLSW model~\cite{Leibovich:1997em}, 
uncertainties in the $B^0$~lifetime~\cite{Amsler:2008zzb}, and the uncertainties in the 
total number of $B^0$~mesons in the data sample.

To calculate these systematic uncertainties, we consider 300 pseudo-experiments
in which one of 15 parameters is randomly varied, using a normal distribution.
The entire analysis chain is repeated for every pseudo-experiment and new fit 
results are obtained, in total for 4500 variations. One standard deviation in 
the pseudo-experiment fit results for a given parameter is used as the systematic
uncertainty in this parameter.

The parameters varied in the pseudo-experiments are as follows:
\begin{enumerate}
\item The corrections on the tracking efficiencies for low momentum
  tracks are 
  varied within their respective uncertainties. To obtain the most
  conservative estimate, the uncertainties in different momentum bins are
  assumed to be fully correlated. Therefore, this component corresponds to
  a single parameter in the toy MC.

\item The lepton identification efficiencies are varied within their
  respective uncertainties~\cite{Hanagaki:2001fz,Abashian:2002bd}.

\item The normalization of the continuum background is not correlated with any
  of the other backgrounds, it is therefore varied individually within the 
  uncertainty on the on- to off-resonance luminosity ratio, which is 1.0\%.

\item
  The normalizations of the remaining five background components are varied within the
  uncertainties listed in Table~\ref{tab:BackgroundFractions}, while taking
  into account the correlations found in the background estimation described 
  in Section~\ref{sec:3c}.

\item
	Uncertainties in the composition of the D** component are accounted for by
	varying each of the components contributing to the $D^{**}$ background within
  the uncertainty reported by the Particle Data Group~\cite{Amsler:2008zzb}. 
  For the resonant modes, this is the uncertainty in the branching fraction products
  $\mathcal{B}(B\to D^{**}\ell\nu)\times\mathcal{B}(D^{**} \to D^{(*)} \pi)$; for the
  non-resonant mode, this is the uncertainty in $\mathcal{B} (B\to D^* \pi \ell \nu)$.
  
\item 
  In addition, the shape of the $q^2$ distributions of the $D^{**}$ components
  is varied according to the LLSW model~\cite{Leibovich:1997em} and the 
  uncertainties on the model parameters as determined in Ref.~\cite{Urquijo:2006wd}.
  
\item  
  The number of $B^0\bar{B}^0$ events is obtained from the product of the number
  of $\Upsilon(4S)$ events in the sample with the branching fraction of $\Upsilon(4S)$
  to a $B^0 \bar{B}^0$ pair. 
  We vary the fraction
  $f_{+-}/f_{00}=\mathcal{B}(\Upsilon(4S)\to B^+B^-)/\mathcal{B}(\Upsilon(4S)
  \to B^0\bar B^0)$ within its uncertainty~\cite{Amsler:2008zzb}. This affects 
  both the overall normalization and the background distributions.

\end{enumerate}

The uncertainties in the reconstruction of the high momentum tracks, the branching ratios
$\mathcal{B}(D^{*+}\to D^0\pi^+)$ and $\mathcal{B}(D^0\to K^-\pi^+)$, 
the number of $\Upsilon(4S)$ events in the sample, and the $B^0$~lifetime affect only 
$\mathcal{F}(1)|V_{cb}|$, not the form factors. Therefore, their uncertainties are
considered by analytical error propagation.

\subsection{Averaging the results of the subsamples}

\begin{figure*}[floatfix]
    \includegraphics[width=0.60\columnwidth]{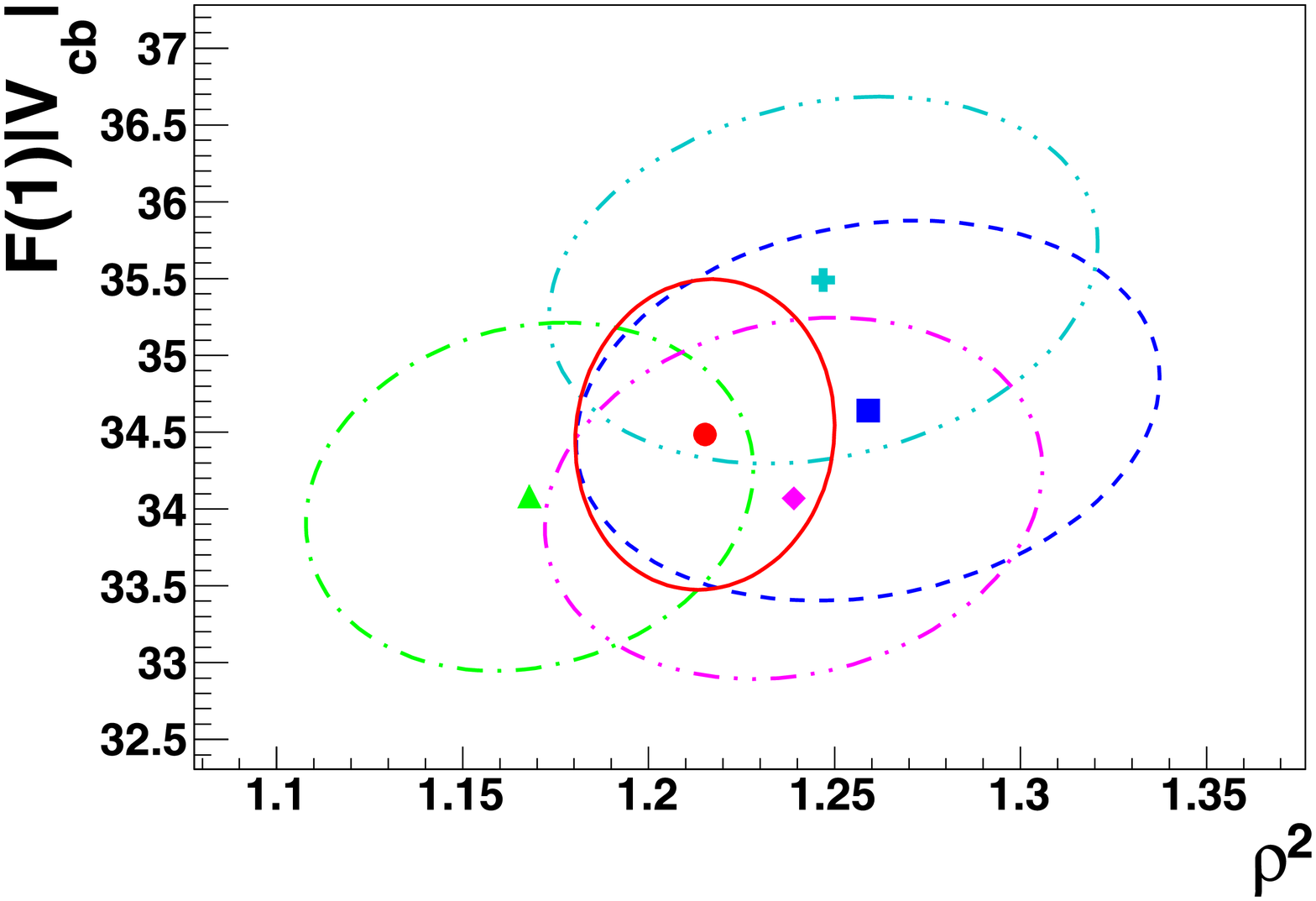}    
    \includegraphics[width=0.60\columnwidth]{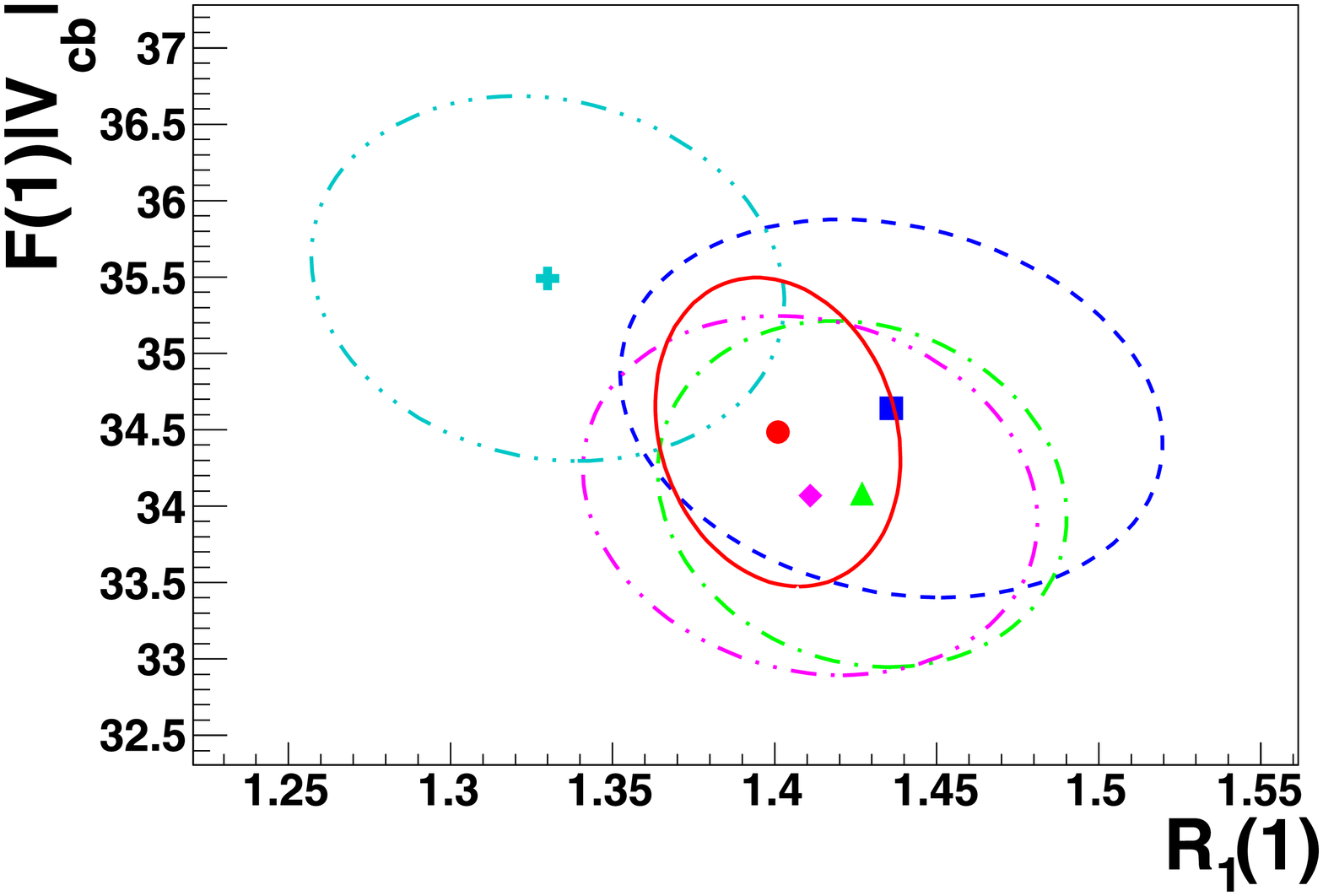}
    \includegraphics[width=0.60\columnwidth]{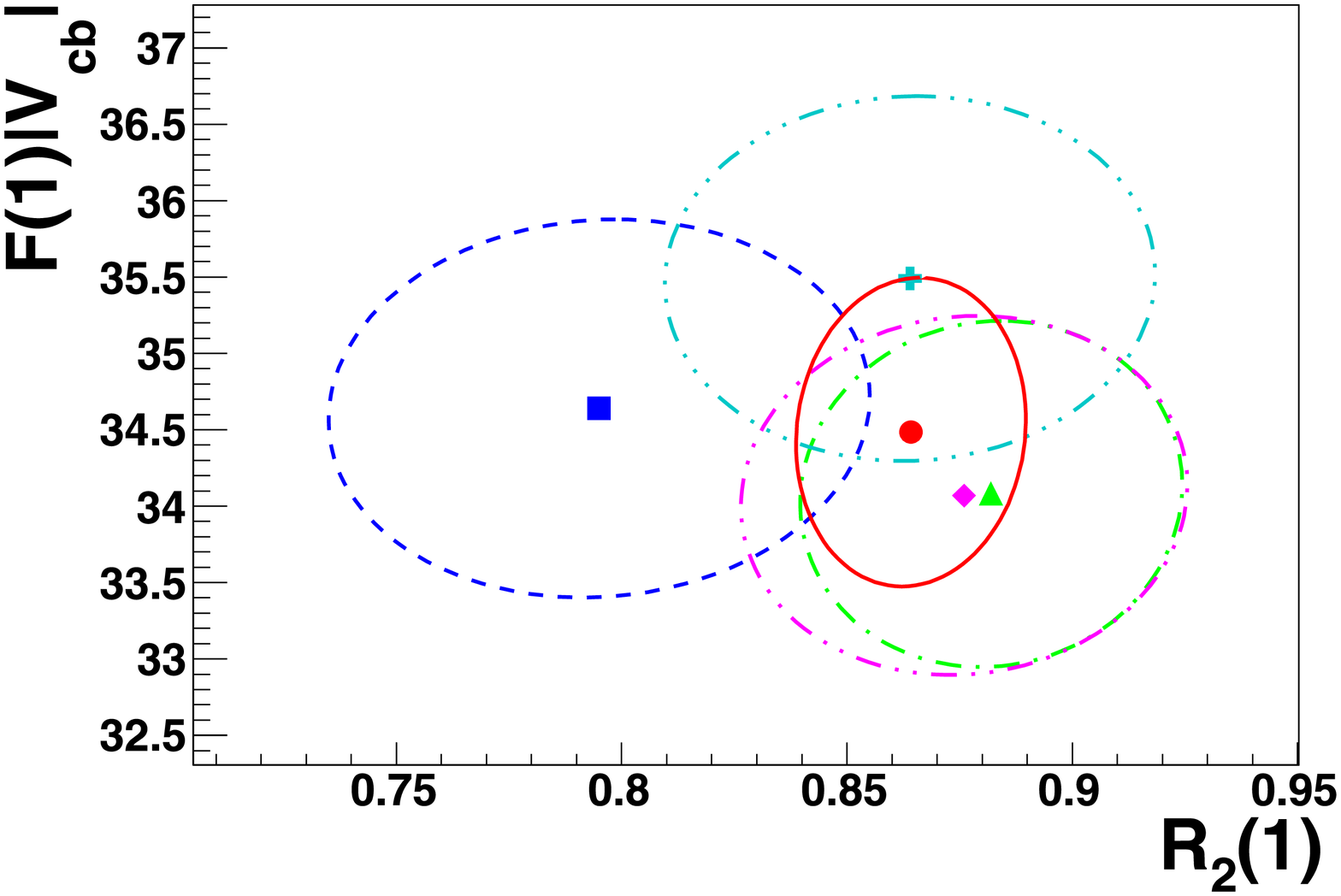}
	\\
    \includegraphics[width=0.60\columnwidth]{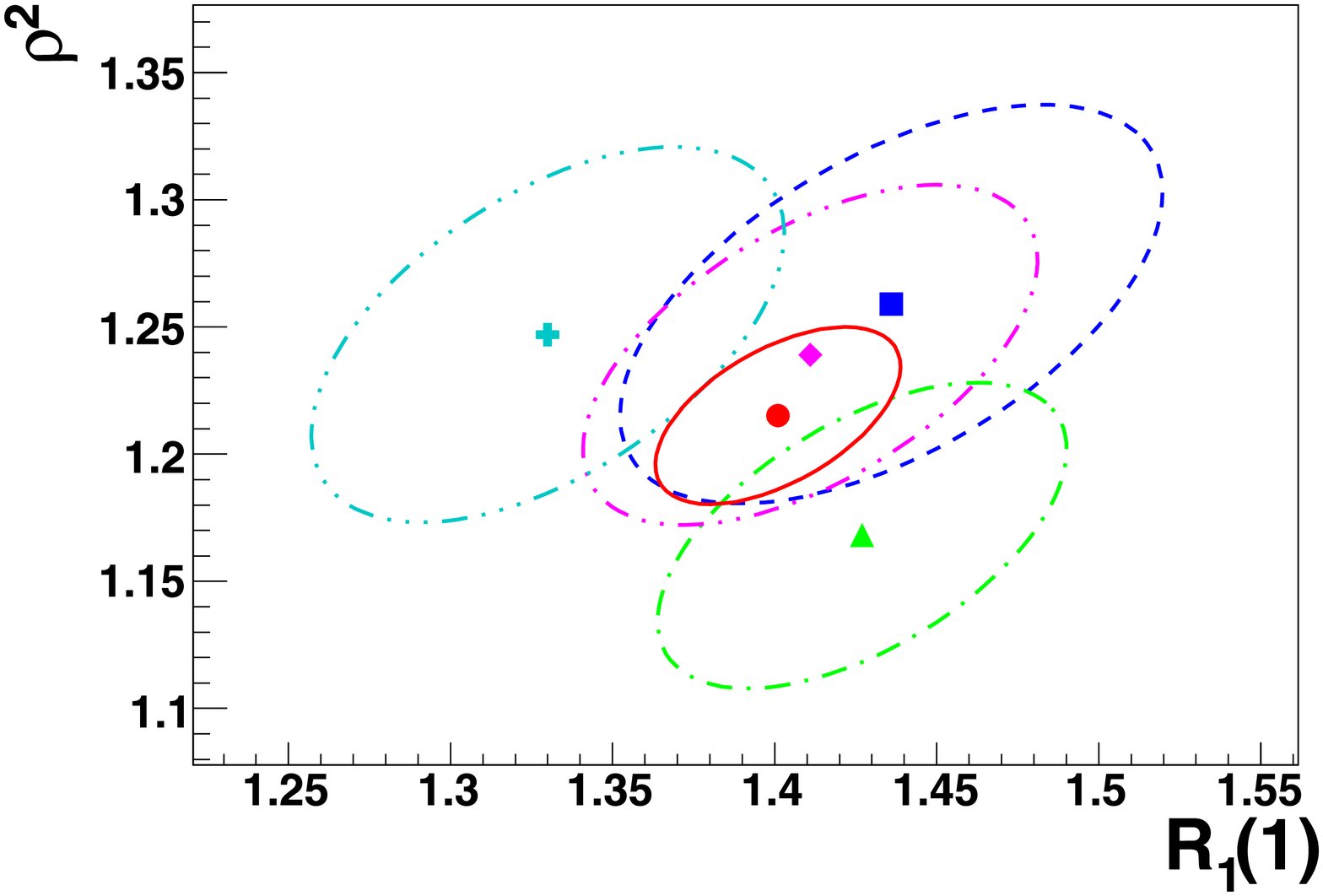}
    \includegraphics[width=0.60\columnwidth]{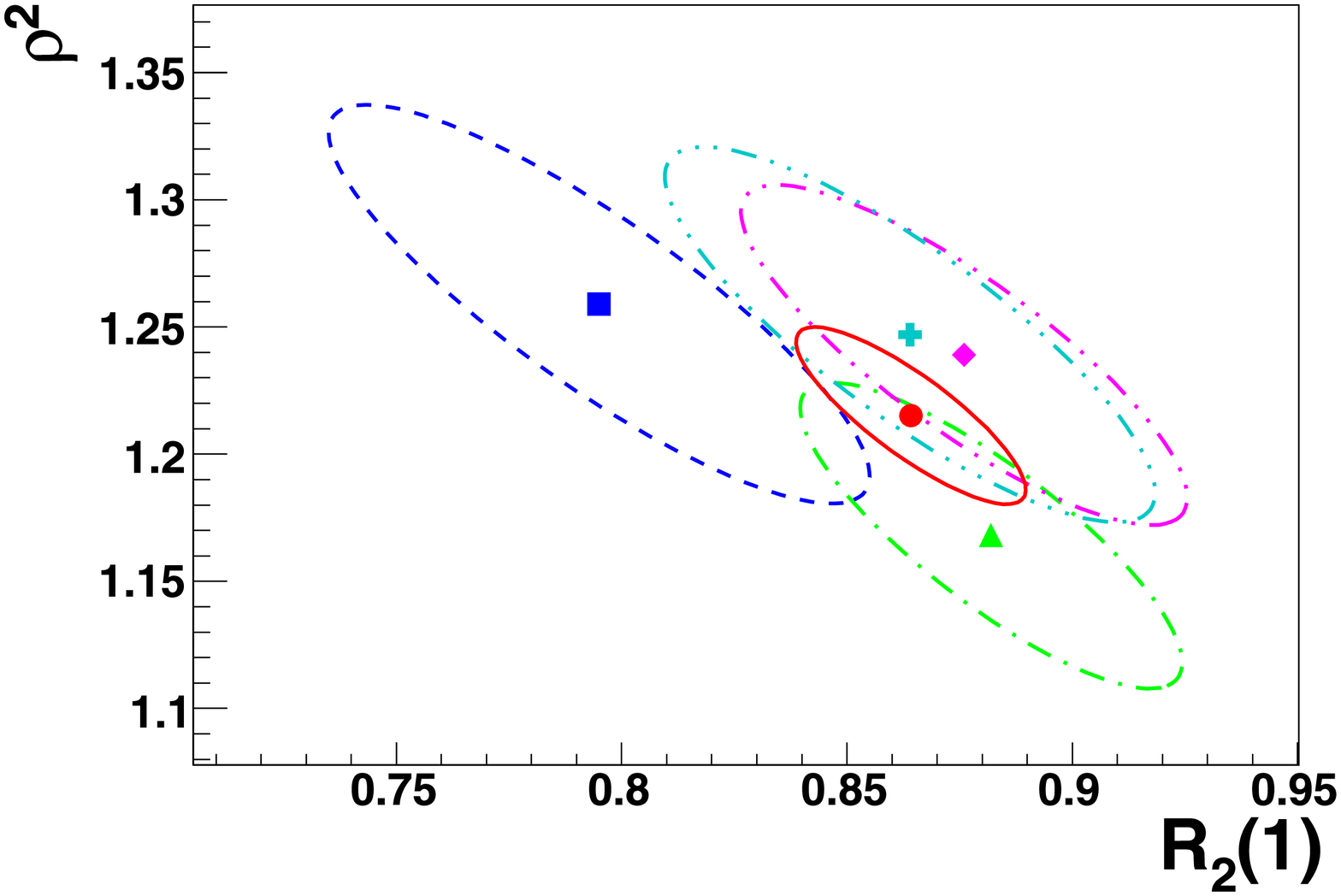}
    \includegraphics[width=0.60\columnwidth]{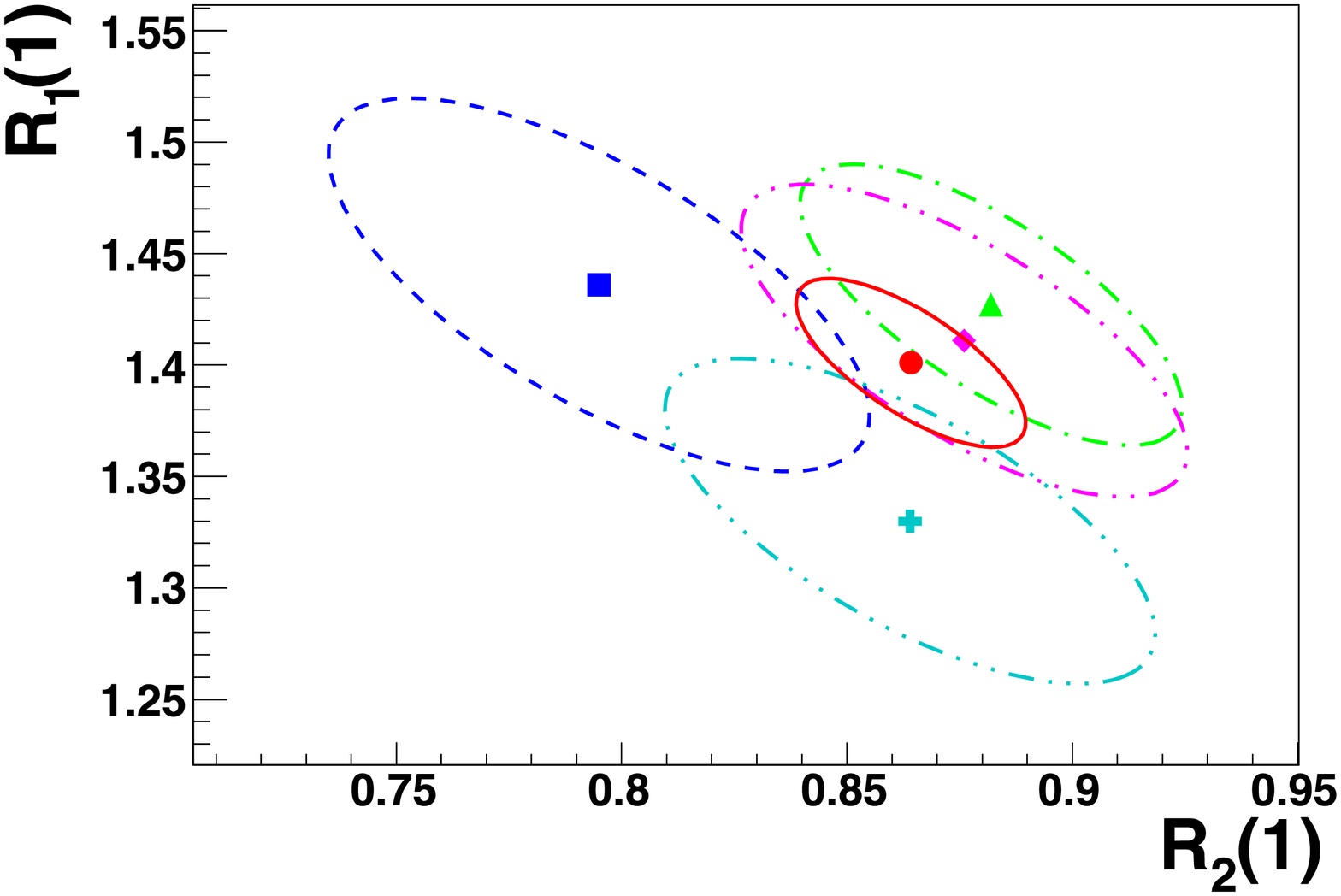}
  \caption{Plots of the result of the averaging procedure. Projections in 
  $\mathcal{F}(1)|V_{cb}|$~vs.~$\rho^2$ (top left), $\mathcal{F}(1)|V_{cb}|$~vs.~$R_1(1)$ (top middle), 
  $\mathcal{F}(1)|V_{cb}|$~vs.~$R_2(1)$ (top right), $\rho^2$~vs.~$R_1(1)$ (bottom left), 
  $\rho^2$~vs.~$R_2(1)$ (bottom middle) and $R_1(1)$~vs.~$R_2(1)$ (bottom right) are shown. The red dot (solid line)
  shows the position (1$\sigma$ ellipse) of the average, the blue
  rectangle (dashed line) the position (1$\sigma$ ellipse) of the
  sub-sample A, the green triangle (dash-dotted line)
  the position (1$\sigma$ ellipse) of the sub-sample B, the magenta diamond
  (dash-double dotted line) the position (1$\sigma$ ellipse) of the 
  sub-sample C and the cyan cross (dash-triple dotted line)
  the position (1$\sigma$ ellipse) of the sub-sample D.} 
  \label{fig:MeteorFig2}
\end{figure*}

To obtain the average of the four sub-samples, which have been measured
independently, we use the algorithm applied by the Heavy Flavor Averaging 
Group~\cite{TheHeavyFlavorAveragingGroup:2010qj} to obtain the world average for 
$|V_{cb}|$ from  semileptonic $B$ decays. This algorithm  combines both the 
statistical and the systematic uncertainties. The correlations of some of these errors
between different samples is considered. For example, the uncertainty on the 
$D^* \to D^0 \pi$ branching fraction will lead to a fully correlated systematic 
uncertainty in each $B\to D^* \ell \nu$ analysis.

The average is obtained with the
{\tt MINUIT} package~\cite{James:1975dr} by using a $\chi^2$ minimization. Here, $N_p$ gives
the total number of fit parameters, in our case $N_p = 4$. When calculating the average 
of  $N$ measurements of the four fit parameters $p_j = \{ \mathcal{F}(1)|V_{cb}|, \rho^2,
R_1(1), R_2(1)\}$, a total of $4\times N$ values are available as inputs, which we label 
as  $V_i = \{ V_1, \dots , V_{4N} \}$. In general this number can be labeled as $N_i$.
Each measurement $V_i$ corresponds to one of the parameters $p_j$, which defines a 
primitive map $\pi(i): i \to p$. The statistical covariance matrix of each measurement 
is known, as well as the correlation between the samples. The latter are 
zero in our case. This information allows one to construct a $4N \times 4N$-dimensional
covariance matrix containing the statistical uncertainties and to obtain the statistical
part of the $\chi^2$ to be minimized:
\begin{equation}
  \chi^2_{\mathrm{stat}} = \sum_i \sum_j \left( V_i - \bar{V}_{\pi(i)} \right) 
	\left( C^{-1} \right)_{ij}
	\left( V_j - \bar{V}_{\pi(j)}\right),
\end{equation}
where $\bar{V}_k$ indicates the average values of the fit parameters.

The systematic uncertainties are implemented by assuming Gaussian error distributions.
The possible bias of input $i$ with respect to the systematic source $s$ can 
therefore be estimated as $ \sigma_{i,s} \times r_s$, where $r_s$ is a normal distributed random
number. Its square is by definition distributed according to a $\chi^2$ distribution with
one degree of freedom, $r_s^2 \sim \chi^2(1)$. The parameters $r_s$ are floating 
fit parameters in {\tt MINUIT}, and in each minimization step the sums
\begin{equation}
	\hat{V}_i = V_i + \sum_s \sigma_{i,s} r_s, \forall i,
\end{equation}
are evaluated. If a systematic uncertainty is associated with two different inputs $i_1$ 
and $i_2$, they are both varied by the same fractional systematic uncertainty at 
the same time. The correlation between systematics is therefore included.

The total $\chi^2$ to be minimized takes the form,
\begin{equation}
  \chi^2 = \sum_i \sum_j \left( \hat{V}_i - \bar{V}_{\pi(i)} \right) 
     \left( C^{-1} \right)_{ij}
     \left( \hat{V}_j - \bar{V}_{\pi(j)}\right) + \sum_s r_s^2,
\end{equation}
and is minimized numerically. The number of degrees of freedom are calculated as
\begin{equation}
  n.d.f = (N_i + N_s) - \underbrace{(N_p + N_s)}_{\mathrm{floated \; parameters}} 
  = N_i - N_p,
\end{equation}
which is the same result one obtains in the case without any systematic uncertainties.
The minimization is numerically stable and yields both the central values and the 
total uncertainties of the full four dimensional average.
\begin{table*}[ht]
	\begin{center}
\begin{tabular}{l|@{\extracolsep{.2cm}}ccccc}
\hline\hline
           &$\rho^2$ & $R_1(1)$ & $R_2(1)$ & $\mathcal{F}(1) |V_{cb} | \times 10^3$ 
           		& $\mathcal{B}( B^0 \rightarrow D^* \ell \nu )$ $[\%]$\\
\hline
Value &     1.214  &     1.401  &     0.864  &     34.6  &     4.58  \\

Statistical Error &     0.034  &     0.034  &     0.024  &      0.2  &     0.03  \\

Systematic Error &      0.009  &     0.018  &     0.008  &      1.0  &     0.26  \\
\hline

Fast track efficiency &       &      &      &     -0.78  &  -0.206    \\

Slow track efficiency & +0.002 & +0.003 & -0.004 & -0.28 & -0.059 \\

$\rho_{\pi_s}$ stability & +0.001 & -0.001 & +0.000 & -0.03 & -0.003 \\

LeptonID &     +0.002  &     +0.006  &     -0.002  &      -0.38  &     -0.100  \\

Norm - $D^{**}$ &     +0.001  &     +0.001  &     -0.001  &     -0.03  &     -0.008  \\

Norm - Signal Corr. &     +0.002  &   -0.003  &   +0.002  &     +0.02  &     +0.006  \\

Norm - Uncorr &     +0.002  &     +0.008  &     -0.003  &      -0.02  &     -0.001  \\

Norm - Fake $\ell$ &     +0.003  &     -0.003 &     -0.001  &    -0.01  &     -0.003 \\

Norm - Fake $D^*$ &     +0.001  &     -0.001  &     +0.000  &      +0.00  &    +0.003 \\

Norm -  Continuum &     +0.002  &     +0.002  &     -0.001  &      +0.00  &    -0.003 \\

$D^{**}$ composition &     +0.004  &     +0.009  &     -0.003  &      -0.10  & -0.025 \\

$D^{**}$ shape &    +0.003  &     +0.005  &     -0.002  &      -0.04  &  -0.011       \\

$N (\Upsilon(4S) )$ &            &            &            &   -0.24  &     -0.063  \\

$f_{+-} / f_{00}$ &     +0.004  &     -0.009  &     +0.003  &   +0.24  &     +0.062  \\

$B^0$ life time &            &            &            &      -0.10  &     -0.027  \\

$\mathcal{B}(D^* \to D^0 \pi_s)$  & &      &            &      -0.13  &     -0.034  \\

$\mathcal{B}(D^0 \to K\pi)$	&      &      &    &      -0.22  &     -0.059  \\

\hline\hline
\end{tabular}

  \end{center}
  \caption{The breakup of the systematic uncertainty in the result
    of the fit to the full sample. The sign + (-) implies whether the fit
  	result moves to larger (smaller) values, if the value of the corresponding
  	systematic parameter is increased.} 
  \label{tab:SystematicBreakdown}
\end{table*}

\begin{table}[floatfix]
	\begin{center}
\begin{tabular}{l|@{\extracolsep{.2cm}}cccc}
\hline\hline

					&         $ \mathcal{F}(1)|V_{cb}|$ &         $\rho^2$ &         $R_1(1)$ &         $R_2(1)$ \\
\hline
$ \mathcal{F}(1)|V_{cb}|$ &     \phantom{-}1.000\phantom{-}  &     \phantom{-}0.625\phantom{-}  &    -0.122\phantom{-}  &    -0.206\phantom{-}  \\

$\rho^2$ &               &     \phantom{-}1.000\phantom{-}  &     \phantom{-}0.575\phantom{-}  &    -0.872\phantom{-}  \\

$R_1(1)$ &               &            &     \phantom{-}1.000\phantom{-}  &    -0.697\phantom{-} \\

$R_2(1)$        &       &            &            &     \phantom{-}1.000\phantom{-}  \\
\hline\hline
\end{tabular}

  \end{center}
  \caption{The statistical correlation coefficients of the four
    parameters in the fit to the full sample.} 
  \label{tab:CorrelationsTotalSample}
\end{table}

\label{sec:finalResult}
Applying this procedure to the four results presented in 
Table~\ref{tab:ParResults1} yields the final
result of this analysis. We obtain 
\begin{eqnarray}
	\mathcal{F}(1)|V_{cb}| &=& (34.5 \pm 0.2 \pm 1.0)\times 10^{-3}, \nonumber\\
	\rho^2 &=& 1.214 \pm 0.034 \pm 0.009, \nonumber\\
	R_1(1) &=& 1.401 \pm 0.034 \pm 0.018, \nonumber\\
	R_2(1) &=& 0.864 \pm 0.024 \pm 0.008,
\end{eqnarray}
with a $\chi^2/n.d.f = 14.3 / 12$ ($P_{\chi^2} = 0.282$). This implies 
excellent agreement between the results, which can also be seen in the projections 
of the minimization, shown in Fig.~\ref{fig:MeteorFig2}.
The corresponding branching fraction for the process $B^0 \to
D^{*-} \ell^+ \nu$ is obtained from the integral of the differential
decay width. We obtain 
\begin{equation}
 \mathcal{B}( B^0 \to D^{*-} \ell^+ \nu) =(4.56 \pm 0.03 \pm 0.26) \%.
\end{equation}

A breakdown of the systematic uncertainties is shown in 
Table~\ref{tab:SystematicBreakdown}. The statistical correlation 
coefficients of the result can be found in 
Table~\ref{tab:CorrelationsTotalSample}.

\section{Model-independent determination of helicity functions}
\label{sec:shapes}
The angular distributions given in Eq.~(\ref{eq:2_1}) are determined by the
kinematic properties of the decay. However, as discussed in section 
\ref{sub:parameterization}, the expressions of the helicity amplitudes and thus the 
distribution in the variable $w$ are based on the parameterization scheme
proposed by Caprini, Lellouch and Neubert~\cite{Caprini:1997mu}.
In this section, we extract the form factor shape of the longitudinal and the
transverse components of Eq.~\ref{eq:2_1} through a fit to the $w$ vs.\ $\cos\theta_V$
distribution. The binning is the same as in the fit
approach described above. The contribution from events with $w>1.5$ is fixed to the
small values predicted by the results of the parameterized fit.

\subsection{Fit procedure}

From Eq.~(\ref{eq:2_1}) we can obtain the double differential decay width
$d \Gamma / dw\,d\cos^{}\theta_V$ by integration over 
$\cos\theta_\ell$ and $\chi$.

If we define
\begin{equation}
	F_\Gamma = \frac{G_F^2\, (m_{B}-m_{D^*})^2 \, m_{D^*}^3 }{4^3 \pi^3}
\end{equation}
and
\begin{eqnarray}
  g^{\pm\pm}(w) &=& 
    \sqrt{w^2-1}(w+1)^2h_{A_1}^2(w)|V_{cb}|^2 \times\nonumber\\
    &&\frac{ 1-2wr-r^2 }{(1-r)^2} 
    \left\{ 1\mp\sqrt{\frac{w-1}{w+1}} R_1(w)\right\}^2,
    \nonumber
  \\
  g^{00}(w) &=&
    \sqrt{w^2-1}(w+1)^2 h_{A_1}^2(w) |V_{cb}|^2 \times\nonumber\\
    &&\left\{1 + \frac{w-1}{1-r} 
    \left( 1- R_2(w) \right) \right\}^2,
\end{eqnarray}
this equation becomes
\begin{eqnarray}
  \frac{d^2\Gamma(B^+\to D^{*0}\ell^+\nu_\ell)}
         {dw\,d\cos^{}\theta_V}
  &=& F_\Gamma\left[\sin^2 \theta_V 
  \left( g^{++} + g^{--} \right) \right.\nonumber\\
  &&\left.+ 2\,\cos^2\theta_V \, g^{00} \right].
  \label{eq:wVsThetaV}
\end{eqnarray}
The quantities $g^{kk}$ correspond to the product of $w$-dependent parts of 
the different helicity combinations and kinematic factors. The one-dimensional
distribution, as given in Eq.~(\ref{eq:2_2}), depends only on the sum of these 
three combinations,
\begin{equation}
  \frac{d\Gamma(B^+\to D^{*0}\ell^+\nu_\ell)}{dw} = 
  \frac{4}{3}F_\Gamma\left(g^{++} + g^{--} + g^{00}\right).
\end{equation}

The bin contents of the two-dimensional histogram in $w$ vs. $\cos\theta_V$ can be obtained 
by integration of Eq.~(\ref{eq:wVsThetaV}) over the corresponding bin area and
considering the reconstruction efficiencies and detector response as described in 
Eq.~\ref{eq:ApplicationOfEfficiencies}. Each bin content can be given as the linear 
combination of two linearly independent parts. The integration of the angular 
distributions is performed analytically, the integration with respect to $w$ defines a
set of dimensionless parameters,
\begin{equation}
  G_{i}^{kk} = \int\limits_{w_i}^{w_{i+1}} \mathrm{d} w \; g^{kk},
\end{equation}
where $w_j = \{ w_1, w_2, \dots, w_{11} \} = \{1, 1.05, \dots, 1.5 \}$ are the bin 
boundaries of the $10$ bins in $w$. In addition, we define $g^{T} = g^{++} + 
g^{--}$, $G_{i}^{T} = G_{i}^{++} + G_{i}^{--}$, $g^{L} = 
g^{00}$ and $G_{i}^{L} = G_{i}^{00}$.

For the $w$ vs.\ $\cos\theta_V$ distribution we the $\chi^2$ function
\begin{equation}
  \tilde{\chi}^2_{k} = \sum_{i=1}^{10}\sum_{j=1}^{10} 
  \frac{N^{2D,\mathrm{obs}}_{k,ij} - N^{2D,\mathrm{exp}}_{k,ij}}{\sigma^2_{N^{2D,\mathrm{exp}}_{k,ij}} },
\end{equation}
for each channel $k$, which depends only on the parameters $G_{i}^{T}$ and 
$G_{i}^{L}$. Here $N^{\mathrm{obs}}$ gives the number of events observed in on-resonance 
data, $N^{\mathrm{exp}}$ the number of expected events, as defined in 
Eq.~\ref{eq:ApplicationOfEfficiencies}, and $\sigma_{N^{\mathrm{exp}}}$ the uncertainty in the 
expected number of events, as given in Eq.~(\ref{eq:sigmaErrorPropagation}). Again we
form the sum of $\chi^2$ functions for each channels and minimize this 
expression numerically using {\tt MINUIT}~\cite{James:1975dr}.

We have tested this fit procedure using generic MC data
samples. All results are consistent with expectations and show no
indication of bias.

\subsection{Results}

\begin{figure*}[floatfix]
  \includegraphics[width=0.8\columnwidth]{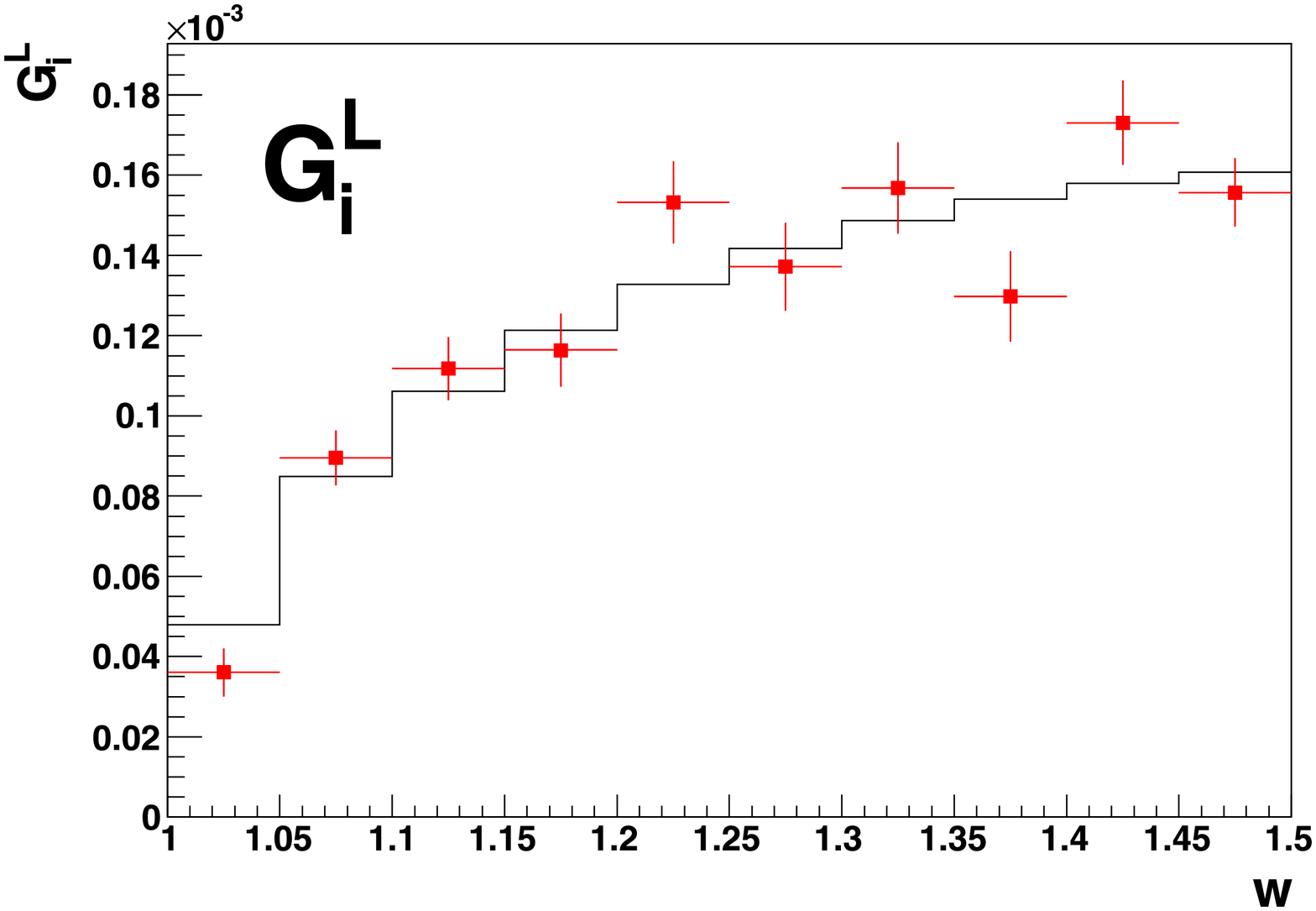}
  \includegraphics[width=0.8\columnwidth]{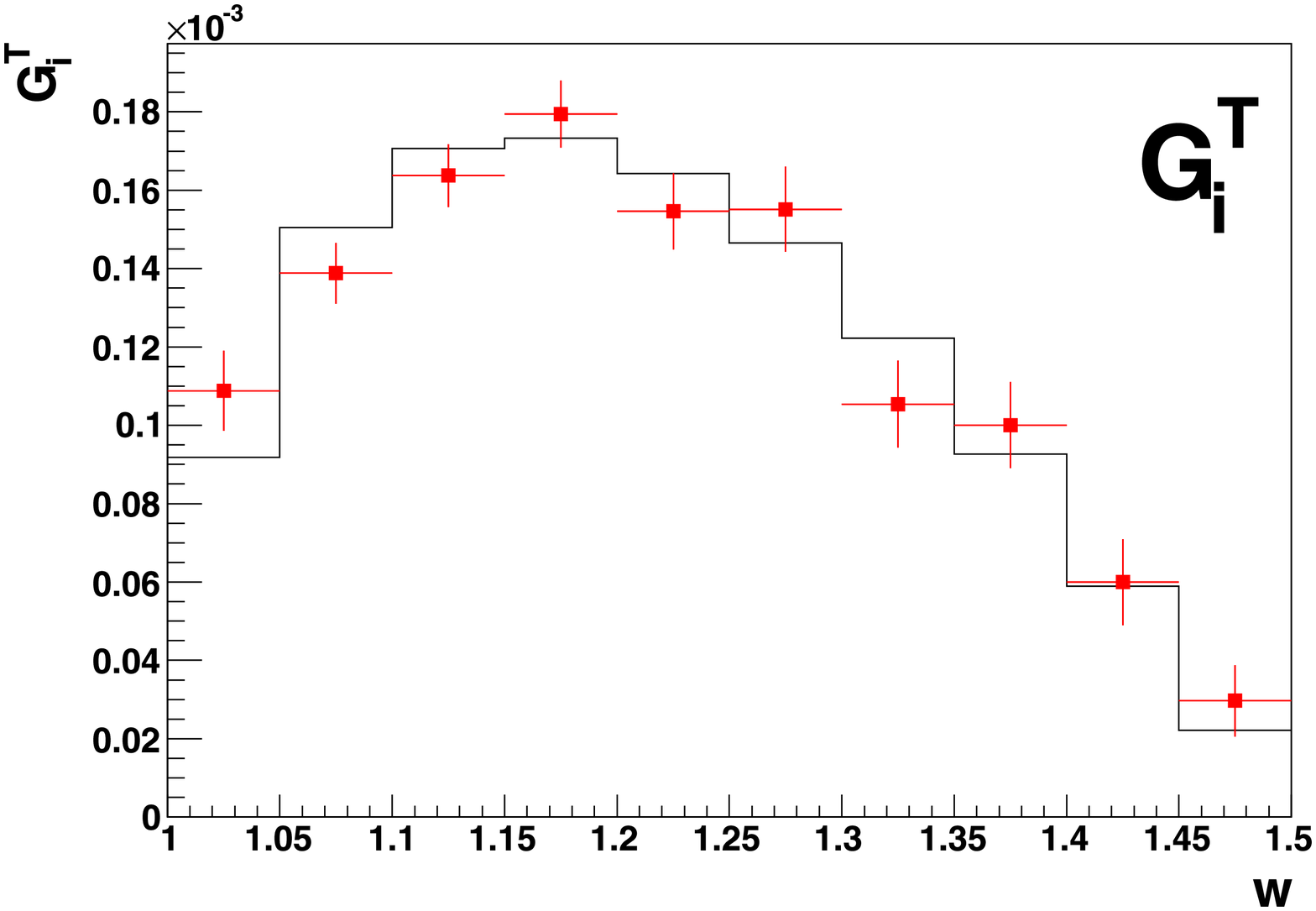}
  \caption{Results of the fit of the helicity functions (red crosses) compared to the 
    prediction obtained by using the parameters obtained by using the parameterization 
    prescription by Caprini, Lellouch and Neubert (solid black line). The left plot 
    shows the results for $G^{L}_i$, the right one for $G^{T}_i$. Only the 
    statistical errors are shown. }
    \label{fig:8}
\end{figure*}

We investigated the largest of the four subsamples -- corresponding to 
about 274 fb$^{-1}$ of data recorded on the $\Upsilon(4S)$ resonance -- to extract 
the helicity shapes. Tables \ref{tab:shapes-results00} and \ref{tab:shapes-resultsT} 
give the results of the fits, where the systematic errors quoted in these tables originate 
from the same sources given in the breakdown in Table~\ref{tab:SystematicBreakdown}.
Many of the dominant systematic uncertainties, such as the track reconstruction or the 
lepton ID uncertainty, are correlated between different bins. The $\chi^2$ of the 
fit is statistically consistent with the number of degrees of freedom, we obtain 
$\chi^2/n.d.f = 175.8 / 179$, $P_{\chi^2} = 55.4 \%$. The results are shown in 
Fig.~\ref{fig:8} and Tables~\ref{tab:shapes-results00}-\ref{tab:shapes-resultsT},
compared to the values obtained using the parameters given in Section~\ref{sec:finalResult}. 

Finally we compute a numerical measure of the agreement between the result obtained 
from this two dimensional fit with the functions predicted by the parameterization of 
Caprini, Lellouch and Neubert~\cite{Caprini:1997mu}. Comparing with the values given
in Table~\ref{tab:ParResults1} is naturally problematic,
since the underlying sample is identical in both fits. Therefore we compare the
extracted shapes to the parameterization using the world average reported by the Heavy Flavor
Averaging Group~\cite{TheHeavyFlavorAveragingGroup:2010qj} in the summer of 2010 and 
form a simple $\chi^2$,
\begin{equation}
	\chi^2 = \sum_i \sum_j (G^{\mathrm{param}}_i - G^{\mathrm{fit}}_i)
	\left( C^{-1}_{\mathrm{stat}} \right)_{ij}
	(G^{\mathrm{param}}_j - G^{\mathrm{fit}}_j)
\end{equation}
where $G^{\mathrm{param}}_i$ are the values obtained using the parameterization by
Caprini, Lellouch and Neubert, $G^{\mathrm{fit}}_i$ are the values obtained by the
unparameterized fit described above and $C_{\mathrm{stat}}$ is the covariance matrix between
the fit parameters, which is also a result of the numerical minimization.
Evaluating this expression yields a $\chi^2 / n.d.f = 29.3 / 20$, 
$P_{\chi^2}=8.3 \%$. This implies satisfactory agreement between the extracted shapes of the
longitudinal and transverse components and the world average parameters.

\begin{table*}
	\begin{center}
\begin{tabular}{l|@{\extracolsep{.2cm}}cc}
\hline\hline

           &  $D^0 \to K\pi, \ell = e$ &  $D^0 \to K\pi, \ell = \mu$ \\
\hline
$G^{T}_1$
&          (     1.187  $\pm$     0.148  $\pm$    0.075  )$\times 10^{-4}$             
&          (     0.982  $\pm$     0.142  $\pm$    0.067  )$\times 10^{-4}$ \\

$G^{T}_2$
&          (     1.514  $\pm$     0.112  $\pm$    0.102  )$\times 10^{-4}$             
&          (     1.239  $\pm$     0.104  $\pm$    0.081  )$\times 10^{-4}$ \\

$G^{T}_3$
&          (     1.594  $\pm$     0.116  $\pm$    0.104  )$\times 10^{-4}$             
&          (     1.685  $\pm$     0.116  $\pm$    0.115  )$\times 10^{-4}$ \\

$G^{T}_4$
&          (     1.809  $\pm$     0.122  $\pm$    0.113  )$\times 10^{-4}$             
&          (     1.760  $\pm$     0.128  $\pm$    0.115  )$\times 10^{-4}$ \\

$G^{T}_5$
&          (     1.649  $\pm$     0.136  $\pm$    0.105  )$\times 10^{-4}$             
&          (     1.484  $\pm$     0.141  $\pm$    0.102  )$\times 10^{-4}$ \\

$G^{T}_6$
&          (     1.511  $\pm$     0.145  $\pm$    0.097  )$\times 10^{-4}$             
&          (     1.572  $\pm$     0.165  $\pm$    0.104  )$\times 10^{-4}$ \\

$G^{T}_7$
&          (     1.135  $\pm$     0.156  $\pm$    0.069  )$\times 10^{-4}$             
&          (     0.974  $\pm$     0.161  $\pm$    0.063  )$\times 10^{-4}$ \\

$G^{T}_8$
&          (     0.933  $\pm$     0.159  $\pm$    0.060  )$\times 10^{-4}$             
&          (     1.072  $\pm$     0.156  $\pm$    0.070  )$\times 10^{-4}$ \\

$G^{T}_9$
&          (     0.631  $\pm$     0.163  $\pm$    0.038  )$\times 10^{-4}$             
&          (     0.571  $\pm$     0.151  $\pm$    0.036  )$\times 10^{-4}$ \\

$G^{T}_{10}$
&          (     0.254  $\pm$     0.141  $\pm$    0.025  )$\times 10^{-4}$             
&          (     0.324  $\pm$     0.122  $\pm$    0.038 )$\times 10^{-4}$ \\
\hline\hline
           & fit to total sample & central value of parametrized fit \\
\hline
$G^{T}_1$
&         (     1.088  $\pm$     0.102  $\pm$     0.069  )$\times 10^{-4}$             
&               0.919$\times 10^{-4}$ \\

$G^{T}_2$
&         (     1.388  $\pm$     0.077  $\pm$     0.092  )$\times 10^{-4}$             
&               1.505$\times 10^{-4}$ \\

$G^{T}_3$
&         (     1.637  $\pm$     0.081  $\pm$     0.108  )$\times 10^{-4}$             
&               1.706$\times 10^{-4}$ \\

$G^{T}_4$
&         (     1.794  $\pm$     0.085  $\pm$     0.113  )$\times 10^{-4}$             
&               1.733$\times 10^{-4}$ \\

$G^{T}_5$
&         (     1.547  $\pm$     0.097  $\pm$     0.101  )$\times 10^{-4}$             
&               1.642$\times 10^{-4}$ \\

$G^{T}_6$
&         (     1.552  $\pm$     0.109  $\pm$     0.100  )$\times 10^{-4}$             
&               1.466$\times 10^{-4}$ \\

$G^{T}_7$
&         (     1.054  $\pm$     0.111  $\pm$     0.065  )$\times 10^{-4}$             
&               1.222$\times 10^{-4}$ \\

$G^{T}_8$
&         (     1.000  $\pm$     0.110  $\pm$     0.064  )$\times 10^{-4}$             
&               0.926$\times 10^{-4}$ \\

$G^{T}_9$
&         (     0.600  $\pm$     0.110  $\pm$     0.035  )$\times 10^{-4}$             
&               0.589$\times 10^{-4}$ \\

$G^{T}_{10}$
&         (     0.297  $\pm$     0.091  $\pm$     0.029  )$\times 10^{-4}$             
&               0.221$\times 10^{-4}$ \\
\hline\hline
\end{tabular}

  \end{center}
  \caption{Results obtained for $G_{i}^{T}$ (dimensionless), compared to the central values obtained 
    from the parameterized fit.} 
  \label{tab:shapes-results00}
	\begin{center}
\begin{tabular}{l|@{\extracolsep{.2cm}}cc}
\hline\hline

           &  $D^0 \to K\pi, \ell = e$ &  $D^0 \to K\pi, \ell = \mu$ \\
\hline
$G^{L}_1$
&          (     0.405  $\pm$     0.083  $\pm$    0.027 )$\times 10^{-4}$             
&          (     0.283  $\pm$     0.090  $\pm$    0.019  )$\times 10^{-4}$ \\

$G^{L}_2$
&          (     0.878  $\pm$     0.096  $\pm$    0.054 )$\times 10^{-4}$             
&          (     0.935  $\pm$     0.099  $\pm$    0.061  )$\times 10^{-4}$ \\

$G^{L}_3$
&          (     1.102  $\pm$     0.109  $\pm$    0.068 )$\times 10^{-4}$             
&          (     1.124  $\pm$     0.112  $\pm$    0.073  )$\times 10^{-4}$ \\

$G^{L}_4$
&          (     1.230  $\pm$     0.128  $\pm$    0.077 )$\times 10^{-4}$             
&          (     1.123  $\pm$     0.133  $\pm$    0.071  )$\times 10^{-4}$ \\

$G^{L}_5$
&          (     1.232  $\pm$     0.137  $\pm$    0.074 )$\times 10^{-4}$             
&          (     1.787  $\pm$     0.151  $\pm$    0.112  )$\times 10^{-4}$ \\

$G^{L}_6$
&          (     1.479  $\pm$     0.149  $\pm$    0.087 )$\times 10^{-4}$             
&          (     1.281  $\pm$     0.159  $\pm$	  0.078  )$\times 10^{-4}$ \\

$G^{L}_7$
&          (     1.426  $\pm$     0.152  $\pm$    0.086 )$\times 10^{-4}$             
&          (     1.727  $\pm$     0.171  $\pm$    0.106  )$\times 10^{-4}$ \\

$G^{L}_8$
&          (     1.458  $\pm$     0.154  $\pm$	  0.083 )$\times 10^{-4}$             
&          (     1.107  $\pm$     0.165  $\pm$    0.067  )$\times 10^{-4}$ \\

$G^{L}_9$
&          (     1.678  $\pm$     0.146  $\pm$    0.100 )$\times 10^{-4}$             
&          (     1.794  $\pm$     0.154  $\pm$    0.111  )$\times 10^{-4}$ \\

$G^{L}_{10}$
&          (     1.592  $\pm$     0.125  $\pm$    0.097 )$\times 10^{-4}$             
&          (     1.527  $\pm$     0.122  $\pm$    0.100  )$\times 10^{-4}$ \\
\hline\hline
           & fit to total sample & central value of parametrized fit \\
\hline           
$G^{L}_1$
&          (     0.361  $\pm$     0.060  $\pm$     0.025  )$\times 10^{-4}$        
&          0.480 $\times 10^{-4}$ \\

$G^{L}_2$
&          (     0.895  $\pm$     0.069  $\pm$     0.056  )$\times 10^{-4}$        
&          0.849$\times 10^{-4}$ \\

$G^{L}_3$
&          (     1.118  $\pm$     0.078  $\pm$     0.070  )$\times 10^{-4}$        
&          1.061$\times 10^{-4}$ \\

$G^{L}_4$
&          (     1.164  $\pm$     0.091  $\pm$     0.073  )$\times 10^{-4}$        
&          1.213$\times 10^{-4}$ \\

$G^{L}_5$
&          (     1.532  $\pm$     0.102  $\pm$     0.094  )$\times 10^{-4}$        
&          1.328$\times 10^{-4}$ \\

$G^{L}_6$
&          (     1.372  $\pm$     0.110  $\pm$     0.082  )$\times 10^{-4}$        
&          1.417$\times 10^{-4}$ \\

$G^{L}_7$
&          (     1.568  $\pm$     0.114  $\pm$     0.095  )$\times 10^{-4}$        
&          1.486$\times 10^{-4}$ \\

$G^{L}_8$
&          (     1.298  $\pm$     0.112  $\pm$     0.076  )$\times 10^{-4}$        
&          1.540$\times 10^{-4}$ \\

$G^{L}_9$
&          (     1.730  $\pm$     0.105  $\pm$     0.104  )$\times 10^{-4}$        
&          1.580$\times 10^{-4}$ \\

$G^{L}_{10}$
&          (     1.557  $\pm$     0.085  $\pm$     0.095  )$\times 10^{-4}$        
&          1.608$\times 10^{-4}$ \\
\hline\hline
\end{tabular}

  \end{center}
  \caption{Results obtained for $G_{i}^{L}$ (dimensionless), compared to the central values obtained 
    from the parameterized fit.} 
  \label{tab:shapes-resultsT}
\end{table*}

\section{Summary and discussion}

Using 711~fb$^{-1}$ of data collected by the Belle experiment, we have
analyzed approximately 120,000 $B^0\to D^{*-}\ell^+\nu_\ell$~decays. A fit to
four kinematic variables fully characterizing these decays yields
measurements of the product of the form factor normalization and $|V_{cb}|$,
$\mathcal{F}(1)|V_{cb}|$,
and of the parameters $\rho^2$, $R_1(1)$ and $R_2(1)$ that enter the
HQET form factor parameterization of this decay. We obtain:
\begin{eqnarray}
	\mathcal{F}(1)|V_{cb}|&=&(34.6\pm 0.2\pm 1.0)\times 10^{-3},  \nonumber\\
	\rho^2&=&1.214\pm0.034\pm 0.009,  \nonumber\\
	R_1(1)&=&1.401\pm 0.034\pm 0.018,  \nonumber\\
	R_2(1)&=&0.864\pm0.024\pm 0.008, \nonumber\\
	\mathcal{B}(B^0 \to D^{*-}\ell^+ \nu_\ell) &=& (4.58 \pm 0.03 \pm 0.26)\%.
	\label{Eq:Finalresults}
\end{eqnarray}
For all these measurements, the first error is the statistical uncertainty and the second is the
systematic uncertainty. 
Using a recent lattice QCD result, $\mathcal{F}(1) = 0.921 \pm 0.013 \pm 0.020$~\cite{Bernard:2008dn}, 
we obtain the following value of $|V_{cb}|$,
\begin{equation}
	|V_{cb}| = 37.5 \pm 0.2 \pm 1.1 \pm 1.0,
\end{equation}
where the third error is due to the theoretical uncertainty on $\mathcal{F}(1)$.
Our results~(\ref{Eq:Finalresults}) are compatible with the recent 
measurements of these quantities by the BaBar experiment~\cite{Aubert:2007rs,
Aubert:2007qs,Aubert:2008yv} as well as with results reported by the ALEPH
~\cite{Buskulic:1996yq}, CLEO~\cite{Briere:2002ew}, DELPHI~\cite{Abreu:2001ic,
Abdallah:2004rz} and OPAL~\cite{Abbiendi:2000hk} experiments. This paper 
supersedes our previous result~\cite{Abe:2001cs}, based on a subset of the 
data used in this analysis. The results presented here give the most precise 
determination of the form factor parameters and $\mathcal{F}(1)|V_{cb}|$ to date.

A direct, model-independent determination of the form factor shapes
has also been carried out and is in good agreement with the HQET-based
form factor parameterization by Caprini {\it et al.}~\cite{Caprini:1997mu}.

\section*{Acknowledgements}

We thank the KEKB group for the excellent operation of the
accelerator, the KEK cryogenics group for the efficient
operation of the solenoid, and the KEK computer group and
the National Institute of Informatics for valuable computing
and SINET3 network support.  We acknowledge support from
the Ministry of Education, Culture, Sports, Science, and
Technology (MEXT) of Japan, the Japan Society for the 
Promotion of Science (JSPS), and the Tau-Lepton Physics 
Research Center of Nagoya University; 
the Australian Research Council and the Australian 
Department of Industry, Innovation, Science and Research;
the National Natural Science Foundation of China under
contract No.~10575109, 10775142, 10875115 and 10825524; 
the Ministry of Education, Youth and Sports of the Czech 
Republic under contract No.~LA10033 and MSM0021620859;
the Department of Science and Technology of India; 
the BK21 and WCU program of the Ministry Education Science and
Technology, National Research Foundation of Korea,
and NSDC of the Korea Institute of Science and Technology Information;
the Polish Ministry of Science and Higher Education;
the Ministry of Education and Science of the Russian
Federation and the Russian Federal Agency for Atomic Energy;
the Slovenian Research Agency;  the Swiss
National Science Foundation; the National Science Council
and the Ministry of Education of Taiwan; and the U.S.\
Department of Energy.
This work is supported by a Grant-in-Aid from MEXT for 
Science Research in a Priority Area (``New Development of 
Flavor Physics''), and from JSPS for Creative Scientific 
Research (``Evolution of Tau-lepton Physics'').


\begin{thebibliography}{99}

\bibitem{Kobayashi:1973fv}
  M.~Kobayashi and T.~Maskawa,
  Prog.\ Theor.\ Phys.\  {\bf 49}, 652 (1973).

\bibitem{Cabibbo:1963yz}
  N.~Cabibbo,
  Phys.\ Rev.\ Lett.\  {\bf 10}, 531 (1963).

\bibitem{Caprini:1997mu}
  I.~Caprini, L.~Lellouch and M.~Neubert,
  Nucl.\ Phys.\  B {\bf 530}, 153 (1998)
  [arXiv:hep-ph/9712417].

\bibitem{ref:0} Throughout this note charge-conjugate decay modes are implied. 

\bibitem{Neubert:1993mb}
  M.~Neubert,
  Phys.\ Rept.\  {\bf 245}, 259 (1994)
  [arXiv:hep-ph/9306320].

\bibitem{Richman:1995wm}
  J.~D.~Richman and P.~R.~Burchat,
  Rev.\ Mod.\ Phys.\  {\bf 67} (1995) 893
  [arXiv:hep-ph/9508250].
  
\bibitem{Amsler:2008zzb}
  K.~Nakamura {\it et al.}  [Particle Data Group],
  J.\ Phys.\ G {\bf 37}, 075021 (2010).

\bibitem{Isgur:1989vq}
  N.~Isgur and M.~B.~Wise,
  Phys.\ Lett.\  B {\bf 232}, 113 (1989).

\bibitem{Isgur:1989ed}
  N.~Isgur and M.~B.~Wise,
  Phys.\ Lett.\  B {\bf 237}, 527 (1990).

\bibitem{Bernard:2008dn}
  C.~Bernard {\it et al.},
  Phys.\ Rev.\  D {\bf 79}, 014506 (2009)
  [arXiv:0808.2519 [hep-lat]].

\bibitem{unknown:2000cg}
  A.~Abashian {\it et al.}  [Belle Collaboration],
  %
  Nucl.\ Instrum.\ Meth.\ A {\bf 479}, 117 (2002).

\bibitem{Kurokawa:2001nw}
  S.~Kurokawa,
  %
  Nucl.\ Instrum.\ Meth.\ A {\bf 499}, 1 (2003), and other papers
  included in this volume.

\bibitem{svd2} Z.Natkaniec {\it et al.} (Belle SVD2 Group), Nucl.\ Instrum.\ and Meth.\ A {\bf 560}, 1(2006).


\bibitem{Lange:2001uf}
  D.~J.~Lange,
  %
  Nucl.\ Instrum.\ Meth.\ A {\bf 462}, 152 (2001).

\bibitem{Brun:1987ma}
  R.~Brun, F.~Bruyant, M.~Maire, A.~C.~McPherson and P.~Zanarini,
  %
  CERN-DD/EE/84-1.
  
\bibitem{Barberio:1993qi}
  E.~Barberio and Z.~Was,
  Comput.\ Phys.\ Commun.\  {\bf 79}, 291 (1994).

\bibitem{Abe:2001hj}
  K.~Abe {\it et al.}  [Belle Collaboration],
  %
  Phys.\ Rev.\ D {\bf 64}, 072001 (2001)
  [hep-ex/0103041].

\bibitem{Fox:1978vu}
  G.~C.~Fox and S.~Wolfram,
  Phys.\ Rev.\ Lett.\  {\bf 41}, 1581 (1978).

\bibitem{Hanagaki:2001fz}
  K.~Hanagaki, H.~Kakuno, H.~Ikeda, T.~Iijima and T.~Tsukamoto,
  Nucl.\ Instrum.\ Meth.\ A {\bf 485}, 490 (2002)
  [hep-ex/0108044].

\bibitem{Abashian:2002bd}
  A.~Abashian {\it et al.},
  Nucl.\ Instrum.\ Meth.\ A {\bf 491}, 69 (2002).

\bibitem{ref:2} Quantities evaluated in the c.m.\ frame are denoted by
  an asterisk.

\bibitem{Leibovich:1997em}
  A.~K.~Leibovich, Z.~Ligeti, I.~W.~Stewart and M.~B.~Wise,
  Phys.\ Rev.\  D {\bf 57} 308 (1998)
  [arXiv:hep-ph/9705467].

\bibitem{Urquijo:2006wd}
  P.~Urquijo {\it et al.}  [Belle Collaboration],
  Phys.\ Rev.\  D {\bf 75} 032001 (2007)
  [arXiv:hep-ex/0610012].

\bibitem{Barlow:1993dm}
  R.~J.~Barlow and C.~Beeston,
  Comput.\ Phys.\ Commun.\  {\bf 77}, 219 (1993).

\bibitem{Brun:1996ro}
	R.~Brun and F.~Rademakers,
  Nucl.\ Inst.\ Meth.\ in\ Phys.\ Res.\ A {\bf 389}, 81-86 (1997).

\bibitem{James:1975dr}
  F.~James and M.~Roos,
  Comput.\ Phys.\ Commun.\  {\bf 10}, 343 (1975).
  
\bibitem{TheHeavyFlavorAveragingGroup:2010qj}
  The Heavy Flavor Averaging Group: D. Asner {\it et al.},
  arXiv:1010.1589 [hep-ex].

\bibitem{Aubert:2007rs}
  B.~Aubert {\it et al.}  [BaBar Collaboration],
  Phys.\ Rev.\  D {\bf 77}, 032002 (2008)
  [arXiv:0705.4008 [hep-ex]].

\bibitem{Aubert:2007qs}
  B.~Aubert {\it et al.}  [BaBar Collaboration],
  Phys.\ Rev.\  Lett {\bf 100}, 231803 (2008)
  [arXiv:0712.3493 [hep-ex]].

\bibitem{Aubert:2008yv}
  B.~Aubert {\it et al.}  [BaBar Collaboration],
  Phys.\ Rev.\  D {\bf 79} 012002 (2009)
  [arXiv:0809.0828 [hep-ex]].

\bibitem{Buskulic:1996yq}
  D.~Buskulic {\it et al.}  [ALEPH Collaboration],
  Phys.\ Lett.\  B {\bf 395} (1997) 373.

\bibitem{Briere:2002ew}
  R.~A.~Briere {\it et al.}  [CLEO Collaboration],
  Phys.\ Rev.\ Lett.\  {\bf 89} 081803 (2002)
  [arXiv:hep-ex/0203032].

\bibitem{Abreu:2001ic}
  P.~Abreu {\it et al.}  [DELPHI Collaboration],
  Phys.\ Lett.\  B {\bf 510} (2001) 55
  [arXiv:hep-ex/0104026].

\bibitem{Abdallah:2004rz}
  J.~Abdallah {\it et al.}  [DELPHI Collaboration],
  Eur.\ Phys.\ J.\  C {\bf 33} (2004) 213
  [arXiv:hep-ex/0401023].

\bibitem{Abbiendi:2000hk}
  G.~Abbiendi {\it et al.}  [OPAL Collaboration],
  Phys.\ Lett.\  B {\bf 482} (2000) 15
  [arXiv:hep-ex/0003013].

\bibitem{Abe:2001cs}
  K.~Abe {\it et al.}  [Belle Collaboration],
  Phys.\ Lett.\  B {\bf 526} (2002) 247
  [arXiv:hep-ex/0111060].

\end{thebibliography}
\end{document}